\documentclass[12pt, draftclsnofoot, onecolumn]{IEEEtran}

\usepackage{amssymb}
\usepackage{amsmath}
\usepackage{cite}
\usepackage{url}
\usepackage{xcolor}
\usepackage{cite,graphicx,amsmath,amssymb}
\usepackage{tabularx,colortbl}
\usepackage[english]{babel}
\usepackage{fancyhdr}
\usepackage{subfigure}
\usepackage{threeparttable}
\usepackage{subfigure}
\usepackage{multirow}
\usepackage{mathrsfs}
\usepackage{balance}
\usepackage{fancyhdr}
\usepackage{mdwmath}
\usepackage{mdwtab}
\usepackage{caption}
\usepackage{amsthm}
\usepackage[xcolor]{changebar}
\usepackage{lipsum}
\usepackage{graphicx}
\usepackage{pifont}
\usepackage{graphicx}
\newcommand{\xmark}{\ding{55}}
\usepackage{setspace}
\usepackage{graphicx}

\usepackage{flafter}
\usepackage{comment}
\usepackage{array}
\usepackage{makecell}

\newtheorem{princ}{Principle}
\newenvironment{pbox} {\begin{princ}}{\hfill \interlinepenalty500 $\Box$\end{princ}}
\newcommand{\bm}[1]{\mbox{\boldmath{$#1$}}}

\newtheorem{theorem}{Theorem}

\newtheorem{lemma}{Lemma}

\newtheorem{corollary}{Corollary}

\makeatletter
\def\ScaleIfNeeded{%
\ifdim\Gin@nat@width>\linewidth \linewidth \else \Gin@nat@width
\fi } \makeatother

\begin{document}

\title{Reconfigurable Intelligent Surfaces:\\
Principles and Opportunities}

\author{
 Yuanwei~Liu,~\IEEEmembership{Senior Member,~IEEE,}
   Xiao~Liu,
     Xidong~Mu,
       Tianwei~Hou,\\
         Jiaqi~Xu,
              Marco Di~Renzo,~\IEEEmembership{Fellow,~IEEE,}
               and Naofal Al-Dhahir~\IEEEmembership{Fellow,~IEEE}

\thanks{

Y. Liu, X. Liu, J. Xu are with the School of Electronic Engineering and Computer Science, Queen Mary University of London, London E1 4NS, UK. (email: yuanwei.liu@qmul.ac.uk; x.liu@qmul.ac.uk; jiaqi.xu@qmul.ac.uk).

X. Mu is with School of Artificial Intelligence and Key Laboratory of Universal Wireless Communications, Ministry of Education, Beijing University of Posts and Telecommunications, Beijing, China (email: muxidong@bupt.edu.cn).

T. Hou is with the School of Electronic and Information Engineering, Beijing Jiaotong University, Beijing 100044, China (email: 16111019@bjtu.edu.cn).

M. Di Renzo is with Universit\'e Paris-Saclay, CNRS, CentraleSup\'elec, Laboratoire des Signaux et Syst\`emes, 3 Rue Joliot-Curie, 91192 Gif-sur-Yvette, France. (emails: marco.di-renzo@universite-paris-saclay.fr).

M. Di Renzo's work was supported in part by the European Commission through the H2020 ARIADNE project under grant agreement number 871464 and through the H2020 RISE-6G project under grant agreement number 101017011.

N. Al-Dhahir is with the Department of Electrical and Computer Engineering, The University of Texas at Dallas, Richardson,
TX 75080 USA. (email: aldhahir@utdallas.edu).

}
}

 \maketitle
\begin{abstract}
Reconfigurable intelligent surfaces (RISs), also known as intelligent reflecting surfaces (IRSs), or large intelligent surfaces (LISs)\footnote[1]{Without loss of generality, we use the name of RIS in the remainder of this paper.}, have received significant attention for their potential to enhance the capacity and coverage of wireless networks by smartly reconfiguring the wireless propagation environment. Therefore, RISs are considered a promising technology for the sixth-generation (6G) of communication networks. In this context, we provide a comprehensive overview of the state-of-the-art on RISs, with focus on their operating principles, performance evaluation, beamforming design and resource management, applications of machine learning to RIS-enhanced wireless networks, as well as the integration of RISs with other emerging technologies. We describe the basic principles of RISs both from physics and communications perspectives, based on which we present performance evaluation of multi-antenna assisted RIS systems. In addition, we systematically survey existing designs for RIS-enhanced wireless networks encompassing performance analysis, information theory, and performance optimization perspectives. Furthermore, we survey existing research contributions that apply machine learning for tackling challenges in dynamic scenarios, such as random fluctuations of wireless channels and user mobility in RIS-enhanced wireless networks. Last but not least, we identify major issues and research opportunities associated with the integration of RISs and other emerging technologies for applications to next-generation networks.

\end{abstract}

\begin{IEEEkeywords}

6G, intelligent reflecting surfaces (IRSs), large intelligent surfaces (LISs), machine learning, performance optimization, reconfigurable intelligent surfaces (RISs), wireless networks
\end{IEEEkeywords}

\section{Introduction}

\begin{figure*}[t!]
    \begin{center}
        \includegraphics[width=18cm]{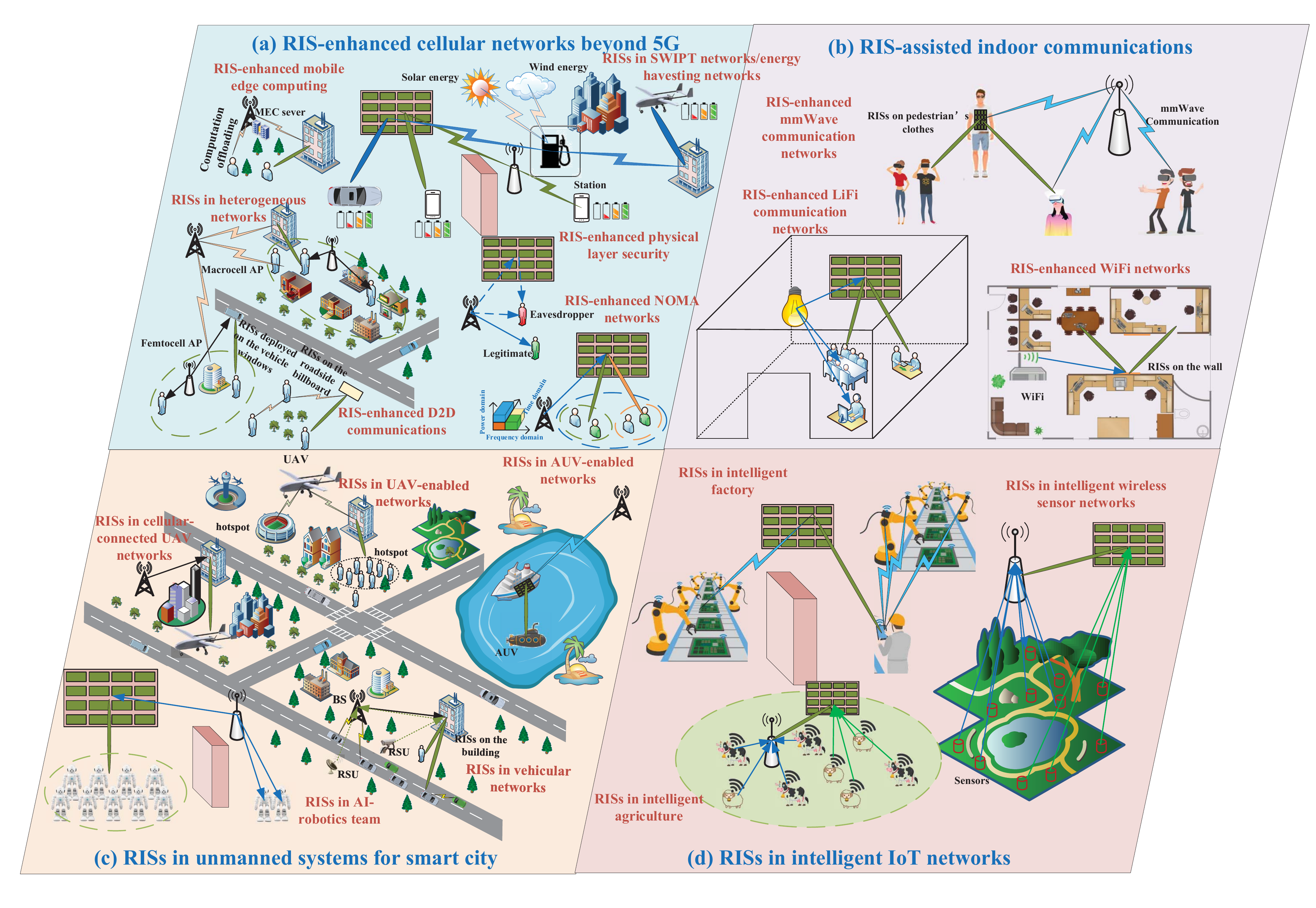}
     \caption{RISs in wireless communication networks.}
            \label{fourparts}
    \end{center}
\end{figure*}

The unprecedented demands for high quality and ubiquitous wireless services impose enormous challenges to existing cellular networks. Applications like rate-centric enhanced mobile broadband (eMBB), ultra-reliable, low latency communications (URLLC), and massive machine-type communications (mMTC) services are the targets for designing the fifth-generation (5G) of communication systems. However, the goals of the sixth-generation (6G) of wireless communication systems are expected to be transformative and revolutionary encompassing applications like data-driven, instantaneous, ultra-massive, and ubiquitous wireless connectivity, as well as connected intelligence~\cite{letaief2019roadmap,saad2019vision}. Therefore, new transmission technologies are needed in order to support these new applications and services. Reconfigurable intelligent surfaces (RISs), also called intelligent reflecting surfaces (IRSs)~\cite{Qingqing2020Towards,cheng2020downlink} or large intelligent surfaces (LISs)~\cite{hou2019mimo,liang2019large}, comprise an array of reflecting elements for reconfiguring the incident signals. Owing to their capability of proactively modifying the wireless communication environment, RISs have become a focal point of research in wireless communications to mitigate a wide range of challenges encountered in diverse wireless networks~\cite{di2020smart,Yang_114}. The advantages of RISs are listed as follows:

\begin{itemize}
\item \textbf{Easy to deploy:} RISs are nearly-passive devices, made of electromagnetic (EM) material. As illustrated in Fig.~\ref{fourparts}, RISs can be deployed on several structures, including but not limited to building facades, indoor walls~\cite{perovic2019channel}, aerial platforms, roadside billboards, highway polls, vehicle windows, as well as pedestrians' clothes due to their low-cost.
\item \textbf{Spectral efficiency enhancement:} RISs are capable of reconfiguring the wireless propagation environment by compensating for the power loss over long distances. Virtual line-of-sight (LoS) links between base stations (BSs) and mobile users can be formed via passively reflecting the impinging radio signals. The throughput enhancement becomes significant when the LoS link between BSs and users is blocked by obstacles, e.g., high-rise buildings. Due to the intelligent deployment and design of RISs, a software-defined wireless environment may be constructed, which, in turn, provides potential enhancements of the received signal-to-interference-plus-noise ratio (SINR).
\item \textbf{Environment friendly:} In contrast to conventional relaying systems, e.g., amplify-and-forward (AF) and decode-and-forward (DF)~\cite{huang2019holographic}, RISs are capable of shaping the incoming signal by controlling the phase shift of each reflecting element instead of employing a power amplifier~\cite{ntontin2019reconfigurable,Bjrnson2019}. Thus, deploying RISs is more energy-efficient and environment friendly than conventional AF and DF systems.
\item \textbf{Compatibility:} RISs support full-duplex (FD) and full-band transmission due to the fact that they only reflect the EM waves. Additionally, RIS-enhanced wireless networks are compatible with the standards and hardware of existing wireless networks~\cite{zhou2020spectral}.
\end{itemize}

Due to the aforementioned attractive characteristics, RISs are recognized as an effective solution for mitigating a wide range of challenges in commercial and civilian applications. There have been many recent studies on RISs and their contributions focus on several application scenarios under different assumptions. As a result, the system models proposed by these research contributions tend to be different. Thus, there is an urgent need to categorize the existing research contributions, which is one of the main goals of this paper.

Fig.~\ref{fourparts} illustrates the applications of RISs in diverse wireless communication networks. In Fig.~\ref{fourparts}(a), RIS-enhanced cellular networks are illustrated, where RISs are deployed for bypassing the obstacles between BSs and users. Thus, the quality of service (QoS) in heterogeneous networks and the latency performance in mobile edge computing (MEC) networks are improved~\cite{cao2019intelligent,bai2019latency}. On the other hand, RISs can act as a signal reflection hub to support massive connectivity via interference mitigation in device-to-device (D2D) communication networks~\cite{cao2020sum}, or RISs can cancel undesired signals by smartly designing the passive beamforming in the context of physical layer security (PLS)~\cite{YangOn2020}. Additionally, RISs can be deployed to strengthen the received signal power of cell-edge users and mitigating the interference from neighbor cells~\cite{pan2019Intelligent}, and the power loss over long distances can be compensated in simultaneous wireless information and power transfer (SWIPT) networks~\cite{Wu_SWIPT_letter,Tang_SWIPT,Pan_SWIPT,Wu_SWIPT}. In Fig.~\ref{fourparts}(b), RIS-assisted indoor communications are illustrated, where RISs can be deployed on walls for enhancing the QoS in some rate-hungry indoor scenarios, such as virtual reality (VR) applications. Additionally, in order to guarantee no blind spots in the coverage area of some block-sensitive scenarios, such as visible light communications~\cite{wang2020performance} and wireless fidelity (WiFi) networks, a concatenated virtual RIS-aided LoS link between the access points (APs) and the users can be formed with the aid of RISs, which indicates that both the propagation links between the APs and the RISs, as well as between the RISs and the users can be in LoS. In Fig.~\ref{fourparts}(c), RIS-enhanced unmanned systems are illustrated. RISs can be leveraged for enhancing the performance of unmanned aerial vehicle (UAV) enabled wireless networks~\cite{li2019reconfigurable}, cellular-connected UAV networks~\cite{ma2019enhancing}, autonomous vehicular networks, autonomous underwater vehicle (AUV) networks, and intelligent robotic networks by fully reaping the aforementioned RIS benefits. For instance, in RIS-enhanced UAV-aided wireless networks, one can adjust the phase shifts of RISs instead of controlling the movement of the UAVs in order to form concatenated virtual LoS links between the UAVs and the users. Therefore, the UAVs can maintain the hovering status only when the concatenated virtual LoS links cannot be formed even with the aid of RISs, which reduces the movement manipulations and the energy consumption of UAVs. In Fig.~\ref{fourparts}(d), RIS-enhanced Internet of Things (IoT) networks are illustrated, where RISs are exploited for assisting intelligent wireless sensor networks~\cite{RIS_fading_to_gamma}, intelligent agriculture, and intelligent factory\cite{Mu_robot}.

There are some short magazine papers~\cite{Liaskos,Qingqing2020Towards,huang2019holographic,gacanin2020wireless}, surveys and tutorials~\cite{di2019smart,liang2019large,basar2019wireless,Gong2020Survey,di2020smart,Wu2020Tutorial} in the literature that introduced RISs and their variants, but the focus of these papers is different from our work. More specifically, Wu~\emph{et al.}~\cite{Qingqing2020Towards} provided an overview of the applications of RISs as reflectors in wireless communications, and identified some challenges and future research opportunities for implementing RIS-assisted wireless networks. Liang~\emph{et al.}~\cite{liang2019large} presented an overview of the reflective radio technology with a particular focus on the large intelligent surface/antennas. In~\cite{di2020smart}, Di Renzo~\emph{et al.} provided a comprehensive overview of employing RISs for realizing smart radio environments in wireless networks, where an electromagnetic-based communication-theoretic framework for analyzing and optimizing metamaterial-based RISs is presented and a survey of recent research contributions on RISs is given. Huang~\emph{et al.}~\cite{huang2019holographic} introduced the concept of holographic multiple-input and multiple-output (MIMO) surfaces (HMIMOS), and discussed both active and passive RISs, encompassing the hardware architectures, operation modes, and applications in communications. In~\cite{Liaskos}, Liaskos~\emph{et al.} presented one kind of RIS prototype, namely the HyperSurface tile, for realizing programmable wireless environments. Gacanin~\emph{et al.}~\cite{gacanin2020wireless} gave an overview of employing artificial intelligence (AI) tools in RIS-assisted radio environments. Di Renzo~\emph{et al.}~\cite{di2019smart} introduced the concept of smart radio environments empowered by RISs, and discussed recent research progresses and future potential challenges. Basar~\emph{et al.}~\cite{basar2019wireless} reviewed recent research efforts on RIS-empowered wireless networks, identified the differences between RISs and other technologies, and presented future research challenges and opportunities. Gong~\emph{et al.}~\cite{Gong2020Survey} surveyed recent research works on RIS-assisted wireless networks and discussed emerging applications and implementation challenges of RISs. From the perspective of enhancing the communication performance, Wu~\emph{et al.}~\cite{Wu2020Tutorial} gave a tutorial on design issues in RIS-assisted wireless networks, including passive beamforming optimization, channel estimation, and deployment design.

Although the aforementioned magazines/surveys/tutorials presented either general concepts or specific aspects of RISs (e.g., from a physics-based or a communication-based perspective), the fundamental performance limits of RISs and some potential applications in wireless networks are not covered. The comparison between widely employed mathematical tools for performance evaluations and optimizations in RIS-enhanced wireless networks is also not discussed. Moreover, a detailed framework based on machine learning (ML) tools for designing RIS-enhanced wireless networks is missing, except for a short magazine paper~\cite{gacanin2020wireless}. Motivated by all the aforementioned considerations, this paper provides a comprehensive discussion of RIS-enhanced wireless network principles, from physics to wireless communications, and discusses research opportunities for exploiting RISs in diverse applications, such as unmanned systems, non-orthogonal multiple access (NOMA), and ML. Table \ref{tab:comparision} illustrates the comparison of this treatise with the existing magazines/surveys/tutorials in the context of RISs.

Against the above observations, our main contributions are as follows.

\begin{table*}[ht!]\scriptsize 
\begin{center}
\resizebox{\textwidth}{!}{\begin{tabular}{|l|l|l|l|l|l|l|l|l|l|l|l|l|l|}\hline
  \multirow{3}{*}{---}   & \multirow{3}{*}{Classifications} & \multirow{3}{*}{Key Contents}  &Wu &Liang &Di Renzo &Huang &Liaskos &Gacanin &Di Renzo   &Basar  & Gong&Wu     &This  \\
                      & &  &\emph{et al.} &\emph{et al.} &\emph{et al.} &\emph{et al.} &\emph{et al.} & \emph{et al.}   & \emph{et al.}   & \emph{et al.} & \emph{et al.} & \emph{et al.}&work\\

                      &  & &\cite{Qingqing2020Towards} &\cite{liang2019large} &\cite{di2020smart} &\cite{huang2019holographic} &\cite{Liaskos} &\cite{gacanin2020wireless} &\cite{di2019smart}     &\cite{basar2019wireless} &\cite{Gong2020Survey}  \cite{Wu2020Tutorial} &\\
     \hline
\multirow{10}{*}{Physics-based} & \multirow{6}{*}{Electromagnetics}  &  Distinguishing ray-optics \& wave-optics perspective & & &$\checkmark$ & & &  &  &    & &    &$\checkmark$ \\
\cline{3-14}
& & Macroscopic description of metasurface-based RIS & & &$\checkmark$ & & &  &$\checkmark$ &$\checkmark$   & &    & \\
\cline{3-14}
& &Surface equivalence theorems & & &$\checkmark$ & & &   &$\checkmark$&  & &   &$\checkmark$ \\
\cline{3-14}
& & Distinguishing far-field \& near-field & & &$\checkmark$ & & &  &  &    & &   &$\checkmark$\\
\cline{3-14}
& & Formulating the reflection coefficient &$\checkmark$ &$\checkmark$ & & & & &  &$\checkmark$    &$\checkmark$ & $\checkmark$  & $\checkmark$\\
\cline{3-14}
& & Power conservation principle & & &$\checkmark$ & & &  & &&&&\\

\cline{2-14}

 & \multirow{2}{*}{RIS Modeling}  &  RIS control mechanism &$\checkmark$ &$\checkmark$ &$\checkmark$ &$\checkmark$ &$\checkmark$ & $\checkmark$ &  $\checkmark$ & $\checkmark$   &$\checkmark$ &  $\checkmark$  & $\checkmark$\\
\cline{3-14}
 & & Typical tunable functions & & &$\checkmark$ & & &  &  $\checkmark$ &  $\checkmark$  &$\checkmark$ &   & $\checkmark$\\

 \cline{2-14}

 & \multirow{2}{*}{Realizations}  &  RIS hardware and prototypes & &$\checkmark$ &$\checkmark$ &$\checkmark$ &$\checkmark$ & &  $\checkmark$ & $\checkmark$   &$\checkmark$ &$\checkmark$    & $\checkmark$\\
\cline{3-14}
 & & RIS synthesis methods & & &$\checkmark$ &$\checkmark$ &$\checkmark$ & &$\checkmark$  &   &&   & $\checkmark$\\

 \cline{1-14}

\multirow{22}{*}{Communication-based}
                   &   & Path loss models  & &$\checkmark$ &$\checkmark$ & & &   &    &$\checkmark$ &$\checkmark$  &$\checkmark$ & $\checkmark$\\
 \cline{3-14}
                   & Performance  & Stochastic geometry  & & &$\checkmark$ & & &  &$\checkmark$   &$\checkmark$ &$\checkmark$ & &$\checkmark$\\
 \cline{3-14}
                   & evaluation & Discussion on stochastic analysis tools & & & & & &  &$\checkmark$   &  &  & &$\checkmark$\\
 \cline{3-14}
                   &  & Discussion on small-scale fading channels & & & & & &  &   &  &  & &$\checkmark$\\
 \cline{2-14}
                    &   & Information-theoretic capacity limits  & & & & & &  &    &    &  & & $\checkmark$\\
 \cline{3-14}
                   &   & Passive beamforming optimization  &$\checkmark$ &$\checkmark$ &$\checkmark$ &$\checkmark$ & &  & &$\checkmark$    & $\checkmark$&$\checkmark$ &$\checkmark$\\
 \cline{3-14}
                   & RIS-aided  & Resource management & & & &$\checkmark$ & &  &   &  &  & &$\checkmark$\\
 \cline{3-14}
                    &communication & Comparison of employed mathematical approaches  & & & & & &  & &  &  &  &$\checkmark$\\
 \cline{3-14}
                    &design & Channel estimation  &$\checkmark$ &$\checkmark$ &$\checkmark$ &$\checkmark$ & &  &   & $\checkmark$ & $\checkmark$ &$\checkmark$  &$\checkmark$\\
 \cline{3-14}
                    & & Deployment design  &$\checkmark$ &$\checkmark$ & & & &  &$\checkmark$  &  &  &$\checkmark$  &$\checkmark$\\
 \cline{3-14}
                    & & Modulation  & & &$\checkmark$ & & & &$\checkmark$   &$\checkmark$  &  &  &\\
 \cline{3-14}
                    & & Localization and sensing  & & &$\checkmark$ & & & &$\checkmark$  &  &  &  &\\

 \cline{2-14}
                     &   & Architecture& & & & & &$\checkmark$    &     &     &  &&  $\checkmark$\\
 \cline{3-14}
                   & ML-empowered & DL& & &$\checkmark$ &$\checkmark$ & & $\checkmark$ &   &$\checkmark$  &$\checkmark$ & &     $\checkmark$\\
   \cline{3-14}
                   & RIS & RL  & & & & & &$\checkmark$  &   &  & & & $\checkmark$\\
   \cline{3-14}
             &  & Comparison of RL-based algorithms & & & & & &  &   &  & & & $\checkmark$\\
   \cline{3-14}
                   &  & Supervised, unsupervised, and federated learning & & & & & &$\checkmark$  &   &  & & &  $\checkmark$\\
   \cline{2-14}
                     &   & NOMA-RIS& & &$\checkmark$ & & &    &     &     &  & &       $\checkmark$\\
 \cline{3-14}
                   &  & PLS-RIS& &$\checkmark$ &$\checkmark$ &$\checkmark$ & &  &   &$\checkmark$  &$\checkmark$ &$\checkmark$  &     $\checkmark$\\
   \cline{3-14}
                   &  & SWIPT-RIS  & &$\checkmark$ &$\checkmark$ &$\checkmark$ & &  &   &  &$\checkmark$ &$\checkmark$ & $\checkmark$\\
   \cline{3-14}
             & Compatibility & UAV-RIS & & &$\checkmark$ & & &  &   & &$\checkmark$ &$\checkmark$ & $\checkmark$\\
   \cline{3-14}
             &  & AV/CV-RIS & & & & & &  &   &  & & & $\checkmark$\\
   \cline{3-14}
             &  & MEC-RIS & & &$\checkmark$ & & &  &   & &$\checkmark$ &$\checkmark$ &$\checkmark$ \\
   \cline{3-14}
             &  & mmWave-RIS & & &$\checkmark$ & & &  & $\checkmark$  &$\checkmark$ &$\checkmark$ &$\checkmark$ &$\checkmark$ \\
   \cline{2-14}
\hline

\end{tabular}}
\end{center}
\caption{Comparison of this work with available magazines/surveys/tutorials. Here, ``DL'' refers to ``Deep Learning'', ``RL'' refers to ``Reinforcement Learning'', ``AV'' refers to ``Autonomous Vehicle'', and ``CV'' refers to ``Connected Vehicle''.}
\label{tab:comparision}
\end{table*}

\begin{itemize}
\item We overview the fundamental principles that govern the operation of RISs and their interaction with the EM signals. We also survey typical RIS functions and their corresponding principles. Specifically, we focus on patch-array based implementation and compare the ray-optics perspective with the wave-optics perspective.
\item We develop performance evaluation techniques for multi-antenna assisted RIS systems. Research contributions are also summarized along with their advantages and limitations.
\item We investigate RISs from the information-theoretic perspective, based on which we review the protocols and approaches for jointly designing beamforming and resource allocation schemes with different optimization objectives. Additionally, the major open research problems are discussed.
\item We discuss the need of amalgamating ML and RISs. After reviewing the most recent research contributions, we propose a novel framework for optimizing RIS-enhanced intelligent wireless networks, where big data analytic and ML are leveraged for optimizing RIS-enhanced wireless networks.
\item We identify major research opportunities associated with the integration of RISs into other emerging technologies and discuss potential solutions.
\end{itemize}

\begin{figure} [t!]
\centering
\includegraphics[width=3.2in]{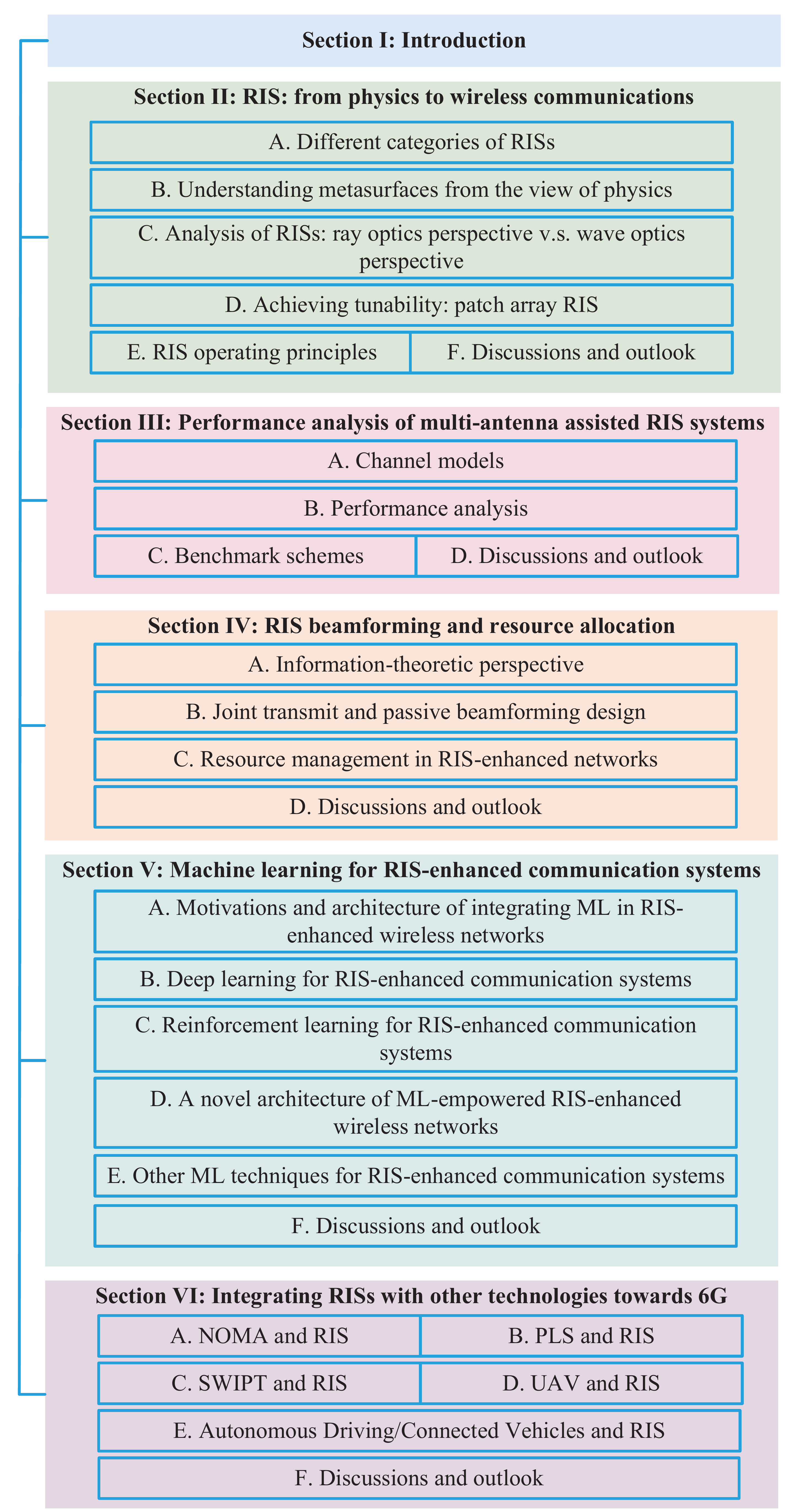}
\caption{Organization of the present paper.}\label{organization}
\end{figure}

As illustrated in Fig.~\ref{organization}, this paper is structured as follows. Section II elaborates on the fundamental operating principles of RIS-enhanced wireless networks. Section III focuses on the performance evaluation of multi-antenna RIS-assisted systems and the main advantages of using RISs in wireless networks. In Section IV, the latest research activities on the joint design of beamforming and resource allocation are discussed. The framework of ML-empowered RIS-enhanced intelligent wireless networks is presented in Section V. Finally, Section VI investigates the integration of RISs with other emerging technologies towards the design and optimization of 6G wireless networks.

\begin{table*}[t!]\scriptsize
\caption{LIST OF ACRONYMS}
\begin{center}
\centering
\begin{tabular}{|l||l|}
\hline
AF & Amplify-and-Forward\\
AO & Alternating Optimization\\
AUV & Autonomous Underwater Vehicle\\
BC & Broadcast Channel\\
BS & Base Station\\
CSI & Channel State Information\\
D2D & Device-to-Device\\
DF & Decode-and-Forward\\
DL& Deep Learning\\
EE & Energy Efficiency\\
EM & Electromagnetic\\
FD & Full-Duplex\\
HD & Half-Duplex\\
IRS & Intelligent Reflecting Surface\\
IoT & Internet of Things\\
LIS & Large Intelligent Surface\\
LoS & Line-of-Sight\\
MEC & Mobile Edge Computing\\
MIMO & Multiple-Input and Multiple-Output\\
MISO & Multiple-Input and Single-Output\\
ML & Machine Learning\\
MOS & Mean Opinion Score\\
NP & Non-deterministic Polynomial-time \\
NOMA & Non-Orthogonal Multiple Access\\
OFDM & Orthogonal Frequency Division Multiplexing \\
OMA & Orthogonal Multiple Access\\
PDF& Probability Density Function \\
PLS & Physical Layer Security\\
QoS& Quality of Service\\
RIS& Reconfigurable Intelligent Surface\\
RL & Reinforcement Learning\\
SCA & Successive Convex Approximation\\
SE & Spectral Efficiency\\
SG & Stochastic Geometry\\
SIC& Successive Interference Cancelation\\
SIMO& Single-Input and Multiple-Output\\
SINR& Signal-to-Interference-plus-Noise Ratio\\
SISO& Single-Input and Single-Output\\
SNR& Signal-Noise Ratio\\
SWIPT& Simultaneous Wireless Information and Power Transfer\\
UAV & Unmanned Aerial Vehicle\\
VLC& Visible Light Communication\\
VR& Virtual Reality\\
WiFi &Wireless Fidelity\\
5G& Fifth-Generation\\
6G& Sixth-Generation\\
\hline
\end{tabular}
\end{center}
\label{table:abbre}
\end{table*}

\section{RIS: From Physics to Wireless Communications}

An RIS is a two-dimensional (2D) material structure with programmable macroscopic physical characteristics. The most important characteristic of an RIS is that its EM wave response can be reconfigured. In contrast to conventional wireless communication networks, the channels between the transmitters and the receivers can be controlled in RIS-aided networks. Thus, the strength of the desired received signal can be enhanced at the terminal devices. In this section, we introduce the fundamental principles which govern the operation of RISs and their interaction with the EM signals. We also survey typical RIS functions and their corresponding principles.

\subsection{Different Categories of RISs}\label{types}
\begin{figure}[t!]
\centering
\includegraphics[width =3.5in]{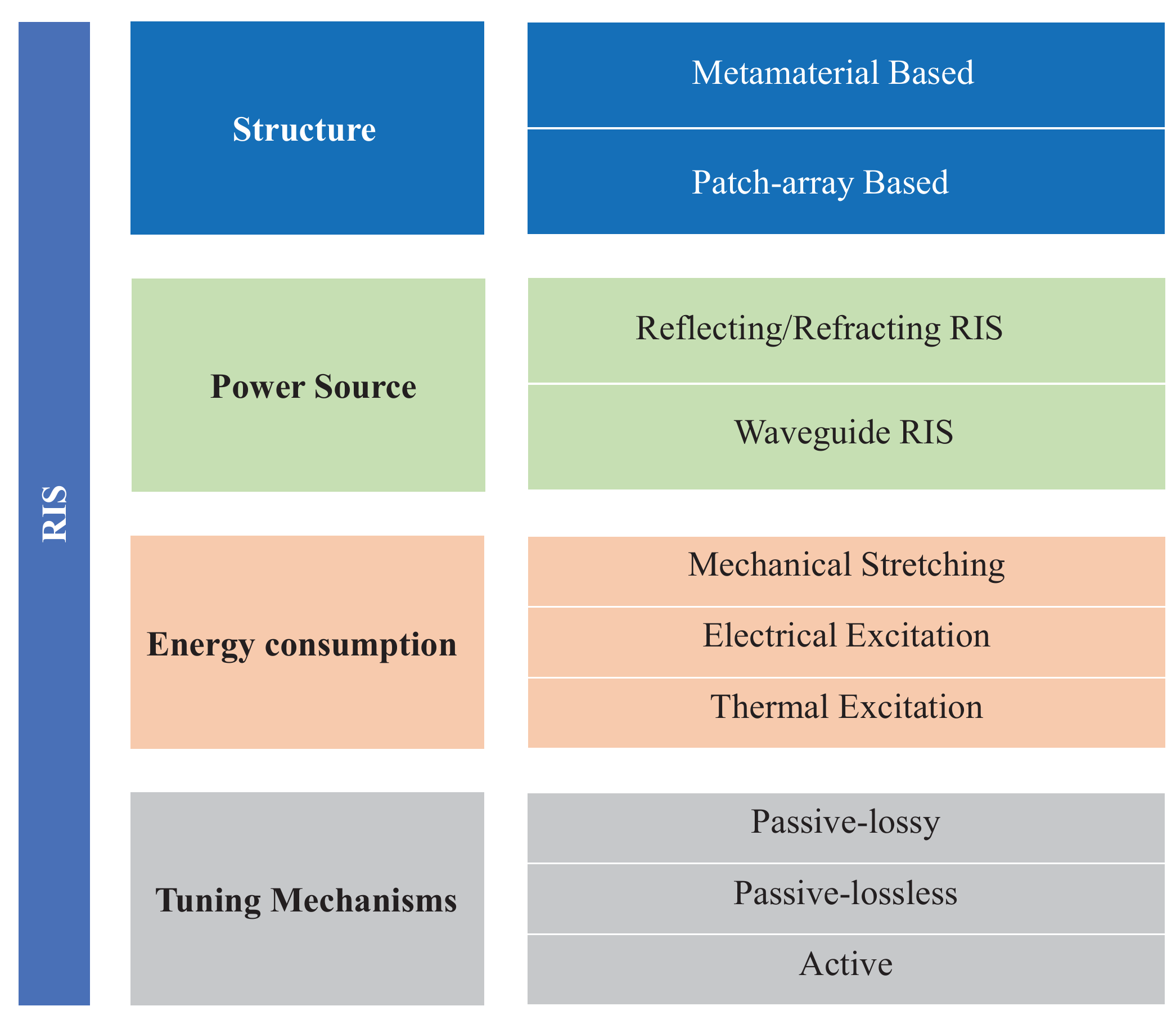}
\caption{Different types of metasurfaces.}
\label{cat}
\end{figure}
Considering their structures, RISs can be realized by using metamaterial or patch-array based technologies. Metamaterial-based RISs are referred to as metasurfaces. Deployed at different locations, RISs can be designed to work as reflecting/refracting surfaces between the BS and the user or waveguide surfaces operating at the BS. Considering the tuning mechanisms, RISs can be reconfigured electrically, mechanically, or thermally. Depending on their energy consumptions, RISs can be categorized as passive-lossy, passive-lossless, or active. The active or passive nature of RISs determines their ultimate performance capabilities. It is worth mentioning, however, that RISs cannot be completely passive because of their inherent property of being configurable. Here, we discuss three important RIS working operations: waveguide~\cite{smith2017analysis}, refraction~\cite{zhu2014dynamic}, and reflection~\cite{dai2018independent}. With the aid of Love's field equivalence principle~\cite{rengarajan2000field}, the reflected and refracted EM field can be studied by introducing equivalent surface electric and magnetic currents~\cite{zhu2014dynamic}. In the three working conditions, the RIS converts and radiates a wave (either induced by an incident wave or fed by a waveguide) into a desired propagating wave in free space. The surface equivalence principles (SEPs) including Love's field equivalence principle and Huygens' principle are introduced in Section \ref{mm}.

\subsubsection{Waveguide RIS}

R. Smith~\emph{et al.}~\cite{smith2017analysis} presented a theoretical study of waveguide-fed metasurfaces. The elements in the metasurface are modeled as uncoupled magnetic dipoles. The magnitude of each dipole element is proportional to the product of the reference wave and each element's polarizability. By tuning the polarizability, the metasurface antenna can perform beamforming. Each element on the metasurface serves as a micro-antenna. Compared to conventional antenna arrays, the compact waveguide metasurface occupies less space and can transmit towards wider angles.

\subsubsection{Refracting RIS}

Viktar S.~\emph{et al.}~\cite{asadchy2016perfect} proposed a theoretical design of a perfectly refracting and reflecting metasurfaces. The authors used an equivalent impedance matrix model so that the tangential field components at the two sides of the metasurface are appropriately optimized. Moreover, three possible device realizations are discussed: self-oscillating teleportation metasurfaces, non-local metasurfaces, and metasurfaces formed by only lossless components. The role of omega-type bianisotropy in the design of lossless-component realizations of perfectly refractive surfaces is discussed.

\subsubsection{Reflecting RIS}

Dai~\emph{et al.}~\cite{dai2018independent} designed a digital coding reflective metasurface. The elements in the metasurface contain varactor diodes with a tunable biasing voltage. By pre-designing several digitized biasing voltage levels, each element can apply discrete phase shifts and achieve beamforming for the reflected wave.

The rest of this section is focused on the operating principles for RISs that operate as reflectors.

\subsection{Understanding Metasurfaces From the View of Physics} \label{mm}

\begin{figure}[t!]
\centering
\includegraphics[width =3.5in]{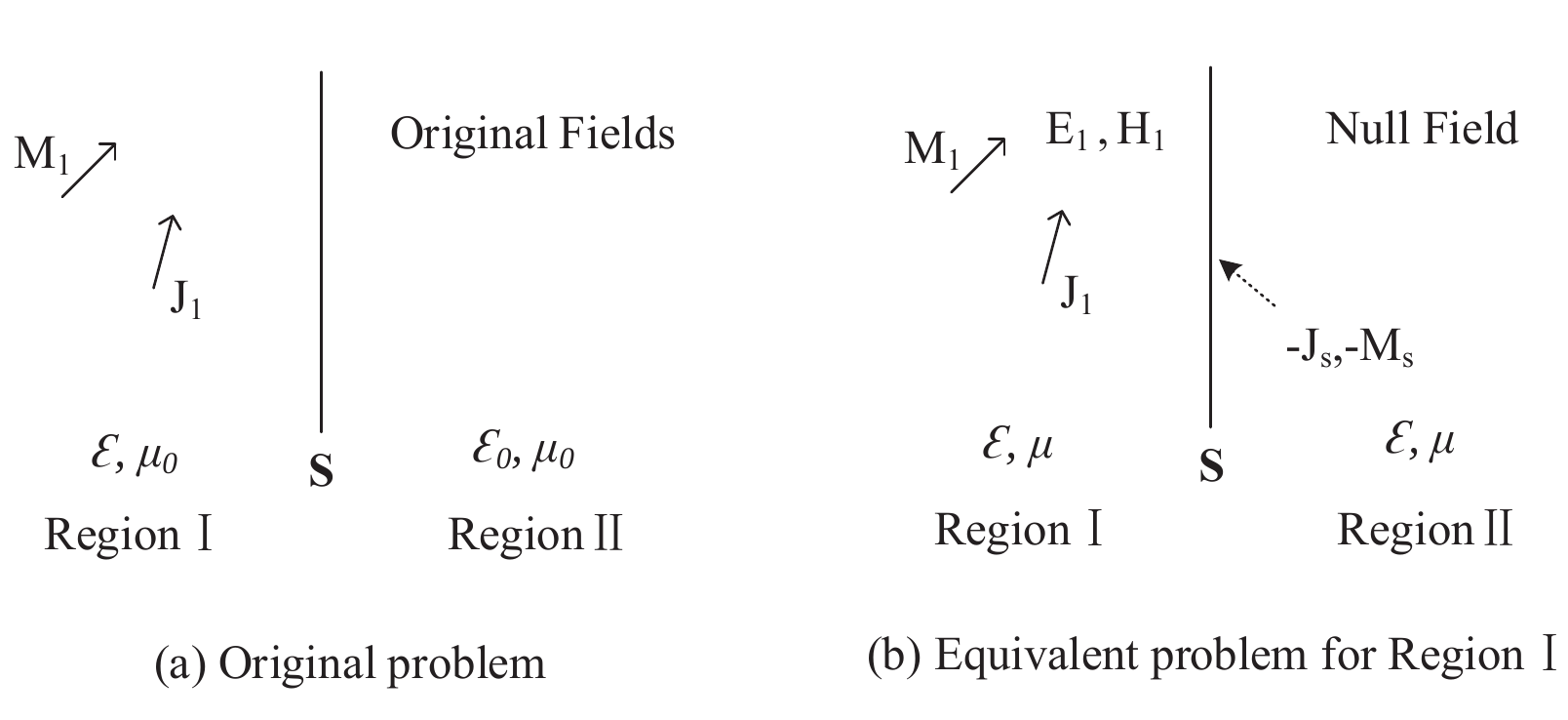}
\caption {Love's equivalence internal problem (for region \textrm{I}). }
\label{love}
\end{figure}

A wireless signal is essentially an EM wave propagating in a three-dimensional space. Attenuation or reduction of the signal strength occurs as the EM wave propagates through the space and interacts with the scattering objects. From basic principles of electromagnetism, the signal power per unit area is proportional to the square of the electric field of the corresponding wave in a given media. As far as reflective and refractive smart surfaces are concerned, this requires the understanding of how the EM waves interact with the surrounding objects. The equivalence principle, especially the SEP, is the building block for studying the EM wave transformations. Some authors also call it Love's field equivalence principle. The principle can be adopted for both external problems (source-free region) and internal problems. Love's field equivalence principle states that the EM field outside or inside a close surface can be uniquely determined by the electric and magnetic currents on the surface. As shown in Fig.~\ref{love}, the equivalent problem for the region \textrm{I} can be reformulated by placing equivalent currents on S that satisfy the boundary condition for each particular case and filling the region \textrm{II} with the same medium of constitutive parameters $\epsilon$ and $\mu$. Thus, the equivalent currents ($-J_s$,$-M_s$), together with the original source currents ($J_1$,$M_1$), radiate the correct fields in region \textrm{I}. The equivalent problem for region \textrm{II} can be formulated similarly.

Love's field equivalence principle is the theoretical foundation for analyzing the radiation pattern of RISs. However, the SEP does not specify how to calculate the EM field produced by the surface currents. To obtain the signal strength at an arbitrary field point, the Huygens-Fresnel principle can be employed. The Huygens-Fresnel principle is a method of analysis applied to problems of wave propagation, which states that every point on a wavefront is itself the source of spherical wavelets, and the secondary wavelets emanating from different points mutually interfere. The sum of these spherical wavelets forms the wavefront. Based on the Huygens-Fresnel principle, the EM field scattered by an RIS (in reflection and refraction) can be quantified analytically.

\begin{figure}[t!]
\centering
\includegraphics[width =3.5in]{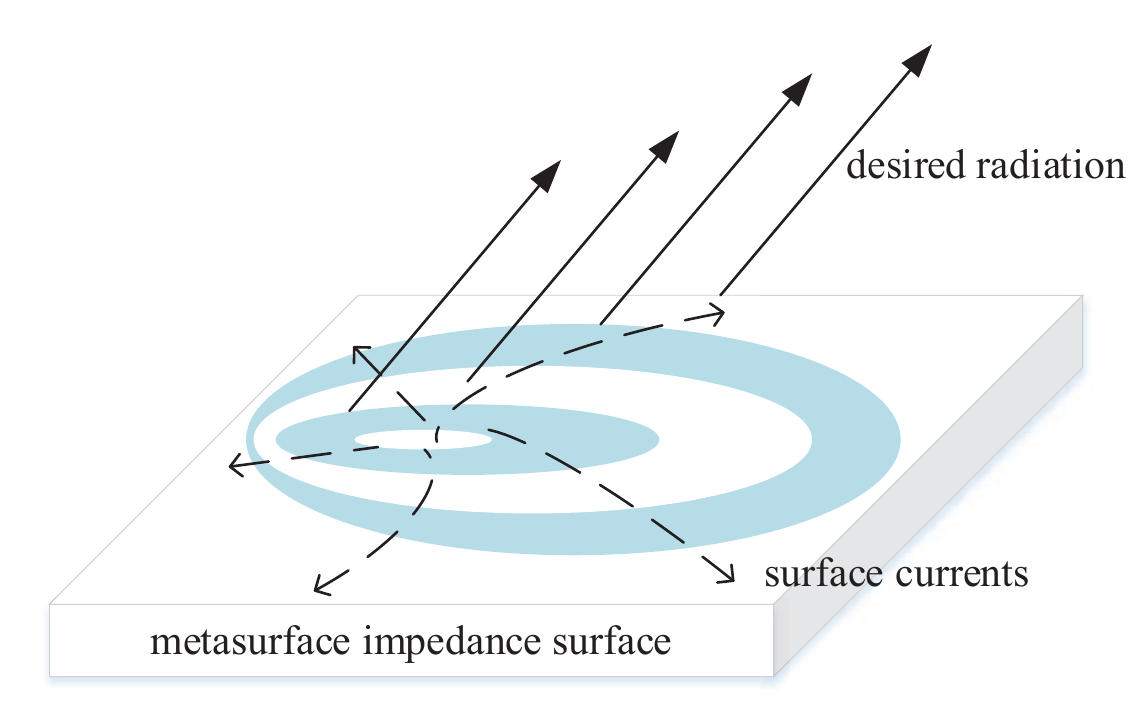}
\caption {Conceptual illustration of the holographic impedance smart surface.}
\label{holo}
\end{figure}

As far as waveguide-based RISs are concerned, the operating principle can be summarized as follows. In \cite{fong2010scalar}, the EM wave manipulation of the waveguide metasurface performs the coupling between three-dimensional free space waves and two-dimensional surface waves. As a result, the metasurface can be regarded as a hologram, which carries additional information about its radiated signal propagating in the 3D space. After being excited by the source, this pre-designed information is coupled into the radiated field. Fig.~\ref{holo} conceptually represents a pre-designed holographic waveguide-based RIS.

\subsection{Analysis of RISs: Ray Optics Perspective v.s. Wave Optics Perspective}\label{design}

\begin{figure}[t!]
\centering
\includegraphics[width =3.5in]{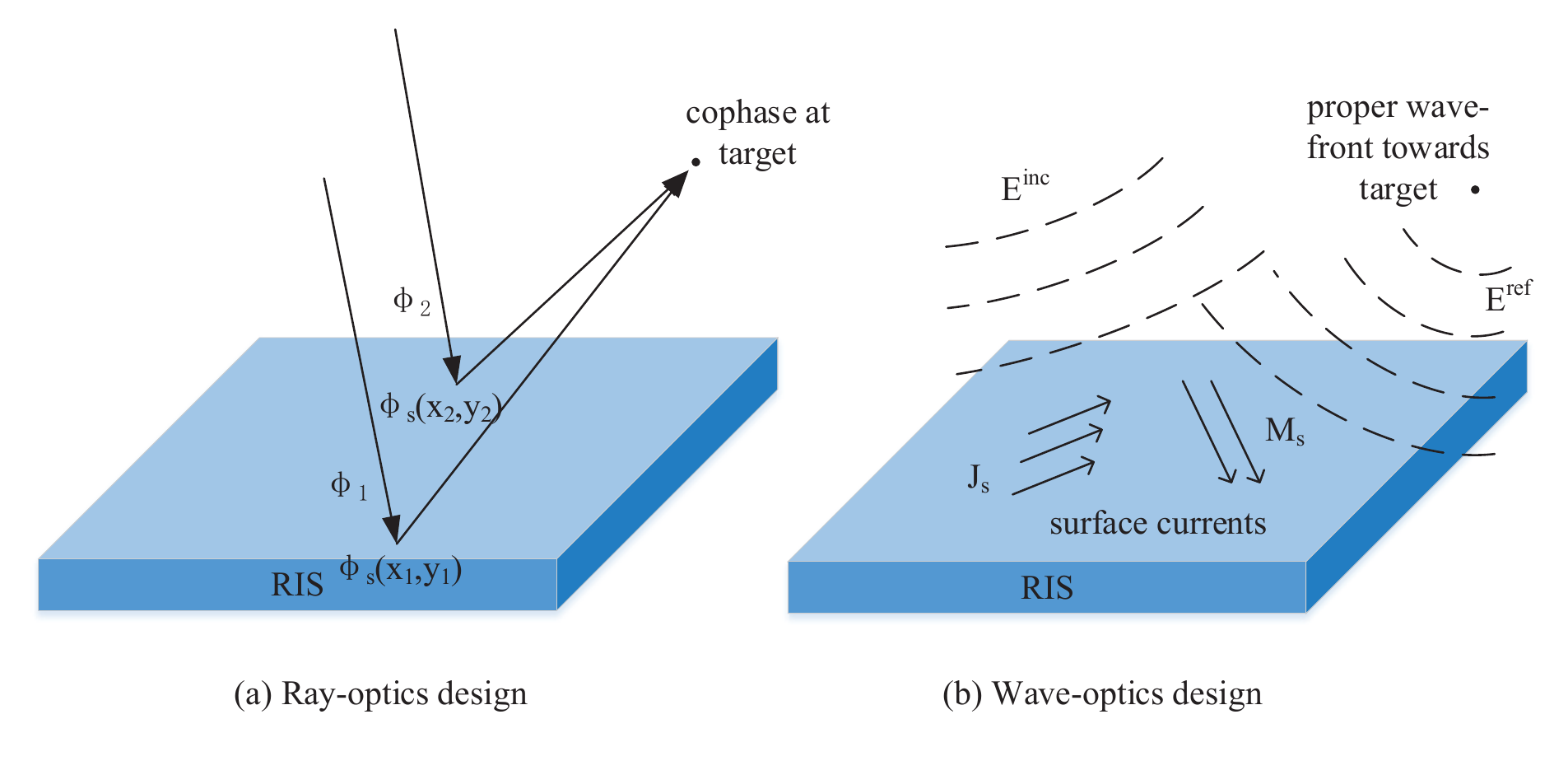}
\caption {Comparison between ray-optics and wave-optics perspective.}
\label{ray}
\end{figure}

To characterize the interaction between an RIS and the impinging EM waves, one can adopt approximations and tools either from the perspective of ray-optics or wave-optics. These two perspectives have been used by physicists for a long time. Even though they are based on some approximations, these two methods of analysis are useful in order to obtain important insights into the interaction of light or radio waves with materials. In the RIS literature, both methods of analysis are often employed. However, the assumptions behind their use and their physical interpretations are intrinsically different. To shed light on their differences and similarities, we compare the two methods in this subsection. As shown in Fig~\ref{ray}(a), from the ray-optics perspective, an EM wave is modeled as a collection of geometrical rays with varying phases. The phase of each ray increases linearly with the optical path length as its traverses through the vacuum or other media. As a result, at each location of the $i$-th ray, a phase (denoted by $\phi_i$) can be defined. When the ray interacts with a material, the phenomenon is studied by determining the relationship between the change of the phase and the material refraction index. The desirable reflected wave is obtained if the ensemble of rays obey the proper co-phase condition. The wave-optics perspective is shown in Fig~\ref{ray}(b), where an EM wave is represented by the corresponding electric field and magnetic field. At each position, each of these two vector fields can be characterized by a time-varying complex-valued vector, with a direction, an amplitude, and a phase. From the wave-optics perspective, the interaction between the wave and the material can be studied using the equivalent principles discussed in the previous subsection. The points with equal phase values form a series of surfaces in space, which we refer to as wavefronts. As a result, the desired scattered waves (reflected or refracted) are obtained if proper wavefront transformations are performed by the RIS.

\begin{table*}[]\scriptsize
\begin{center}
\centering
\begin{tabular}{|l|l|l|}
\hline
                                 & \textbf{Ray-optics} & \textbf{Wave-optics} \\ \hline
Wave representation              & Geometrical rays    & Vector fields        \\ \hline
Theoretical foundation           & Snell's law         & Maxwell's equations    \\ \hline
Surface profile                  & Phase discontinuity & Surface impedance    \\ \hline
Requirement of the reflected wave & Co-phase condition  & Proper wavefront     \\ \hline
Power flow                       & Not accurate       & Accurate    \\ \hline
\end{tabular}
\caption{Comparing different wave representation perspectives}
\label{table:pers}
\end{center}
\end{table*}

\subsubsection{Comparing Ray-Optics and Wave-Optics Perspectives}

Table \ref{table:pers} highlights some of the differences between the two methods of analysis. Compared with wave-optics, the ray-optics perspective is a stronger simplification of the real system. As a result, it is easier to be adopted and can produce a quick prediction about the RIS design. However, ray-optics methods fail when considering the RIS power flow. Wave-optics methods, on the other hand, predict the power flow by using the Poynting vector, which enables us to study the local and overall RIS power consumption. This is an important issue to consider when designing and manufacturing RISs. For example, the authors of~\cite{asadchy2016perfect} and~\cite{estakhri2016wave} point out that it is impossible to realize lossless plane-wave beam steering with locally passive RISs. One has to adopt the wave-optics perspective to study the power flow of the system. Moreover, the authors of~\cite{estakhri2016wave} state that, according to their simulation results, one can expect increasingly improved performance if the RISs are designed based on wave-optics approximation in comparison to those designed based on the ray-optics approximation. In conclusion, both perspectives have their advantages and limitations. However, adopting the wave-optics perspective is the most appropriate choice for most cases. A study of the differences, in terms of power flow and reflection coefficient, between ray optics and wave optics methods of analysis is reported in \cite{di2020smart}.

\subsection{Achieving Tunability: Patch Array RIS}\label{tune2}

The EM characteristics of an RIS, such as the phase discontinuity, can be reconfigured by tuning the surface impedance, through various mechanisms. Apart from electrical voltage, other mechanisms can be applied, including thermal excitation, optical pump, and physical stretching. Among them, electrical control is the most convenient choice, since the electrical voltage is easier to be quantized and controlled by field-programmable gate array (FPGA) chips. The choice of RIS materials include semiconductors~\cite{zhu2013active} and graphene~\cite{emani2015graphene}.

\begin{figure}[t!]
\centering
\includegraphics[width =3.5in]{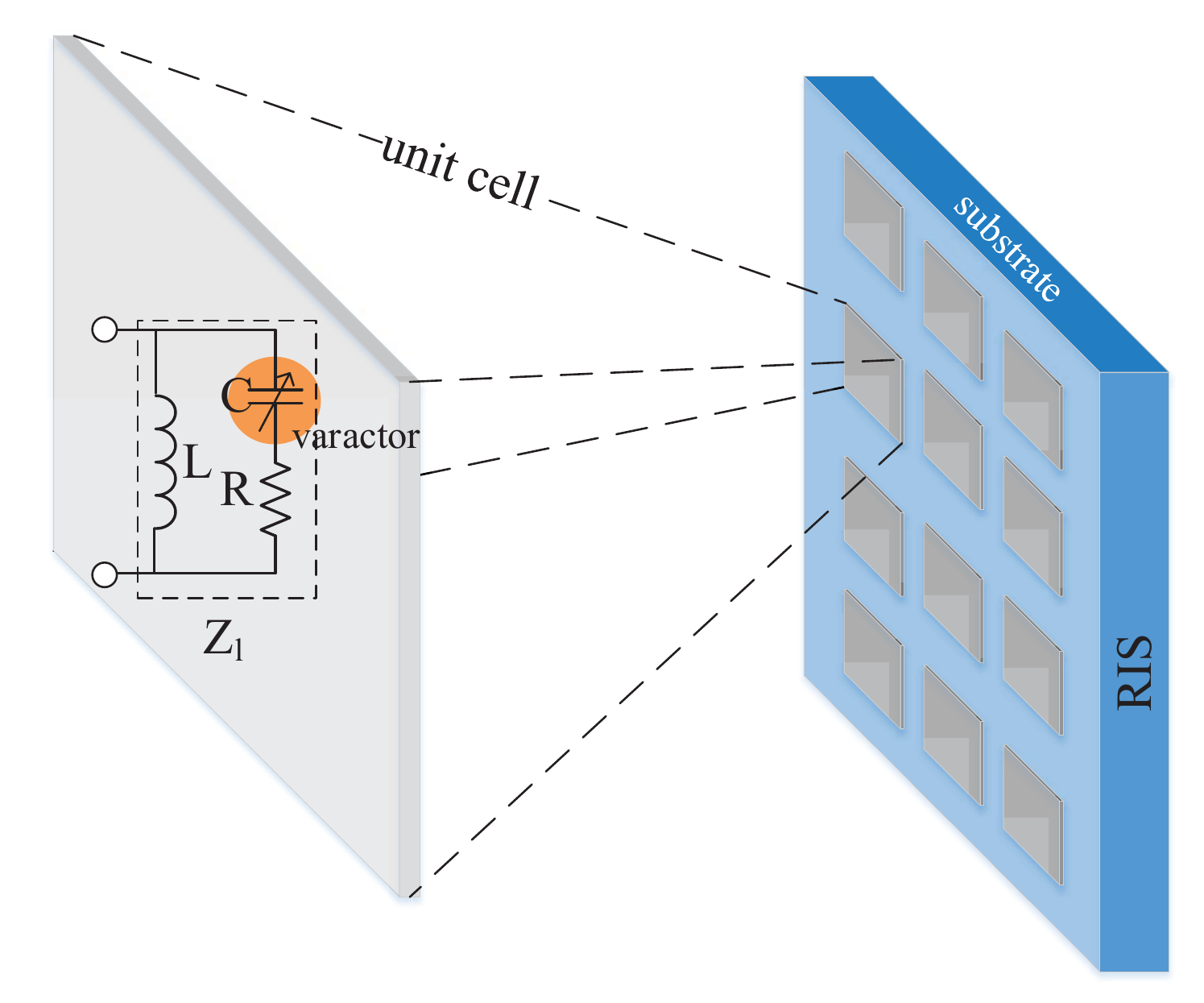}
\caption{Schemetic diagram of the varactor RIS.}
\label{vara}
\end{figure}

Regardless of the tuning mechanisms, we focus our attention on patch-array smart surfaces in the following text. The general geometry layout of this type of RIS can be modeled as a periodic (or quasi-periodic in the most general case) collection of unit cells integrated on a substrate. For ease of description, we limit our discussion to RISs that are based on a local design, in which the cells do not interact with each other. A local design usually results in the design of sub-optimal RISs. A comprehensive discussion about local and non-local designs can be found in \cite{di2020smart}. To characterize the tunability of the RIS, the method of equivalent lumped-element circuits can be adopted. As shown in Fig. \ref{vara}, the unit cell is equivalent to a lumped-element circuit with a load impedance $Z_l$. Particularly, the equivalent load impedance can be tuned by changing the bias voltage of the varactor diode. When modeling patch-array RISs in wireless communication systems, we can characterize, under a local design, each of its unit cells through an equivalent reflection coefficient. For example, the reflection coefficient of the $i$-th cell can be modeled as follows:
\begin{equation}\label{refl}
    r_i=\beta_i \cdot e^{j\phi_i}
\end{equation}
where $\beta_i$ and $\phi_i$ correspond to the amplitude response and the phase response, respectively. As shown in~\cite{gradoni2020end}, the equivalent reflection coefficient depends on the tuning impedance of the lumped circuit that controls each unit cell, as well as the self and mutual impedances (if mutual coupling cannot be ignored) at the ports of the RIS. In particular, as shown in~\cite{abeywickrama2019intelligent} and~\cite{gradoni2020end}, $\beta_i$ and $\phi_i$ in \eqref{refl} are usually not completely independent with each other, i.e., $\beta_i=f(\phi_i)$. In the following sections, $\Phi(\vec{r_x})$ refers to the phase discontinuity introduced by the RIS as a function of the position on the RIS, and $\phi_{mn}$ refers to the phase discontinuity of the $(m,n)$-th element of a patch-array RIS.

Existing designs of patch-array RISs can apply discrete phase control and, in some cases, amplitude control. Arun~\emph{et al.}~\cite{arun2020rfocus} designed \textit{RFocus}, which is a two-dimensional surface with a rectangular array of passive antennas. The size of each passive unit is $\lambda/4 \times \lambda/10$ and the EM waves are either reflected or refracted. The authors show that the \textit{RFocus} surface can be manufactured at a low cost, and that it can improve the median signal strength by $9.5$ times. Welkie~\emph{et al.}~\cite{welkie2017programmable} developed a low-cost device embedded in the walls of a building to passively reflect or actively transmit radio signals. Dunna~\emph{et al.}~\cite{dunna2020scattermimo} realized \textit{ScatterMIMO}, which uses a smart surface to increase the scattering in the environment. In their hardware design, each reflector unit uses a patch antenna connected to four open-ended transmission lines. The transmission lines provide $0$, $\pi/2$, $\pi$ or $3\pi/2$ phase shifts. Based on measurements, it is shown that \textit{ScatterMIMO} increases the throughput by factor of two and the signal-noise ratio (SNR) by 4.5 dB as compared with baselines schemes. It is worth mentioning that tunability can be achieved with implementations other than patch-array based surfaces. For example, PIVOTAL COMMWARE proposed a new technique called \textit{Holographic Beam Forming (HBF)}. The proposed holographic beamformer has a low cost and power consumption, as compared with other transmission technologies, such as massive MIMO and phased arrays.

\subsection{RIS Operating Principles}\label{prin}

Considering single-beam reflection, a patch-array based RIS can be configured to serve a terminal device in the far-field and near-field regions. Among the many operating functions and configurations of RISs, \textit{anomalous reflection} and \textit{beamforming} are widely used in the context of wireless communications. Adopting the wave-optics perspective, anomalous reflection is a wavefront transformation from a plane wave to another plane wave, while beamforming is a wavefront transformation from a plane wave to a desired wavefront. Adopting the ray-optics perspective, we present the operating principles of these two configurations in the following text. As far as anomalous reflection is concerned, the RIS is designed to reflect an incident beam to a far-field terminal, following the generalized laws of reflection~\cite{bell1969generalized}. As far as beamforming (also called focusing) is concerned, the incident wave is focused towards a targeted region, often referred to as the \textit{focal point}. The required RIS configuration follows the co-phase condition~\cite{huang2008reflectarray}. The relation between these two operating principles is discussed in detail in~\cite{Renzo_cylindrical_mirror_scatter}. Before presenting these two different principles, the physical distinction between the near-field region and the far-field region is clarified.

\subsubsection{Near Field v.s. Far Field}

In the spirit of dimensional analysis, the characteristics of a system can be represented by dimensionless numbers. In order to separate the near-field region from the far-field region, a proper dimensionless number is needed. Let $L$ and $R_F$ denote the antenna aperture size of the RIS and the focal distance, respectively. Assume that $z$ is the distance of a particular field point to the RIS. Theoretically, the far-field and near-field regimes can be differentiated as follows: The distance of $2L^2/\lambda$ is a commonly used criterion to decide the boundary between the near-field and far-field regions (see ~\cite{johnson1973determination}, equation (1)). The position corresponding to $z=2L^2/\lambda$ is the boundary between the near-field region and the far-field region. This result comes from the inspection of the power density variation with the distance between a field point and the RIS. Within the near field where $z<2L^2/\lambda$, the power density shows significant variations. The peak position of the power density in the near-field region, namely $R_F$, changes with different RIS configurations. Using proper co-phase conditions, beam focusing can be achieved within the near-field of the RIS. It is worth mentioning that, in general, the boundary between the near-field and far-field regimes depends on the specific configuration of the RIS, as it was recently remarked in~\cite{Renzo_cylindrical_mirror_scatter}.

In general, the essential difference between the near-field region and the far-field region is how the power density changes with distance. Consider, for example, that the RIS focuses the wave within an area $a$. The total energy incident on the RIS is proportional to the solid angle, $\Omega$ spanned by the surface area of the RIS with respect to the location of the transmitter. After the reflection, the transmitted energy is spread over the area $a$. Thus, the power density around the focal point is proportional to $\Omega /a$. Moreover, according to \cite{arun2020rfocus}, the area $a$ is proportional to $\lambda^2 (1+4z^2/L^2)$, as a result of the Abbe diffraction limit. In the far-field region, the second term inside the brackets dominates and $\Omega /a$ is proportional to $L^2\Omega /z^2$, which is the typical spherical dissipation of the signal power with the distance. In the near-field region, the first term dominates and the area $a$ becomes very small. As a result, a high focusing gain can be achieved.

With the aid of an RIS, in general, the signal can be enhanced in both the near-field and far-field regimes. However, the rationale of this enhancement is different. For near-field applications, the RIS is expected to enhance the signal strength for users located at targeted locations with respect to the RIS, while reducing the signal at other locations. For far-field applications, the RIS is typically expected to enhance the signal strength for users located at targeted angles with respect to the RIS.

\begin{figure}[t!]
\centering
\includegraphics[width =3.5in]{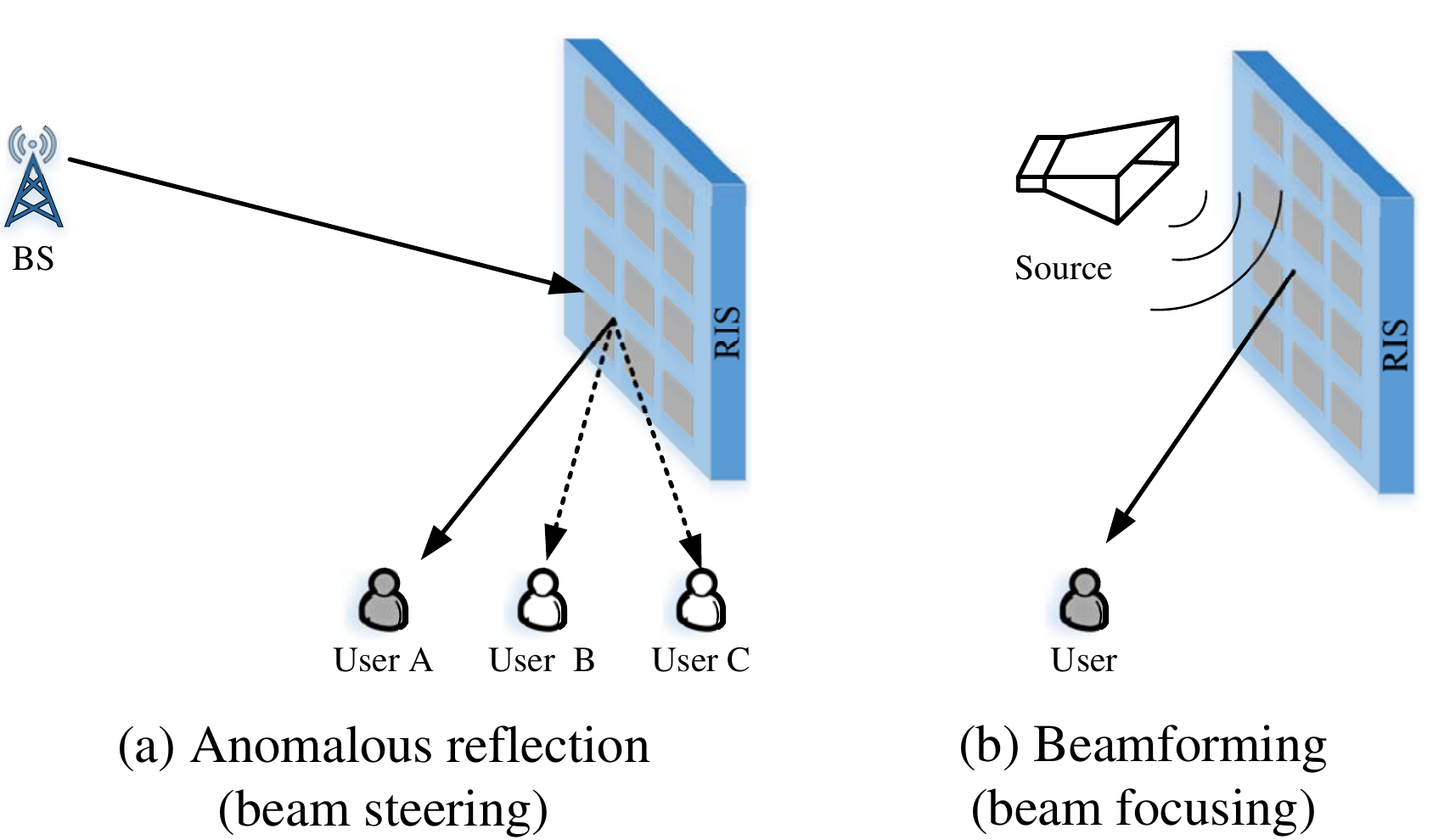}
\caption{Typical functions of reflecting surfaces.}
\label{func}
\end{figure}

In the following text, adopting the ray-optics perspective, we discuss the generalized laws of refraction and reflection, as well as the corresponding co-phase condition.

\subsubsection{The Generalized Laws of Refraction and Reflection}

From a geometrical optics perspective, anomalous reflection and refraction from an RIS can be described by using the generalized laws of refraction and reflection~\cite{bell1969generalized}, which is a natural derivation of both Fermat's principle and the boundary conditions governed by Maxwell's equations.

\begin{figure}[t!]
\centering
\includegraphics[width =3.5in]{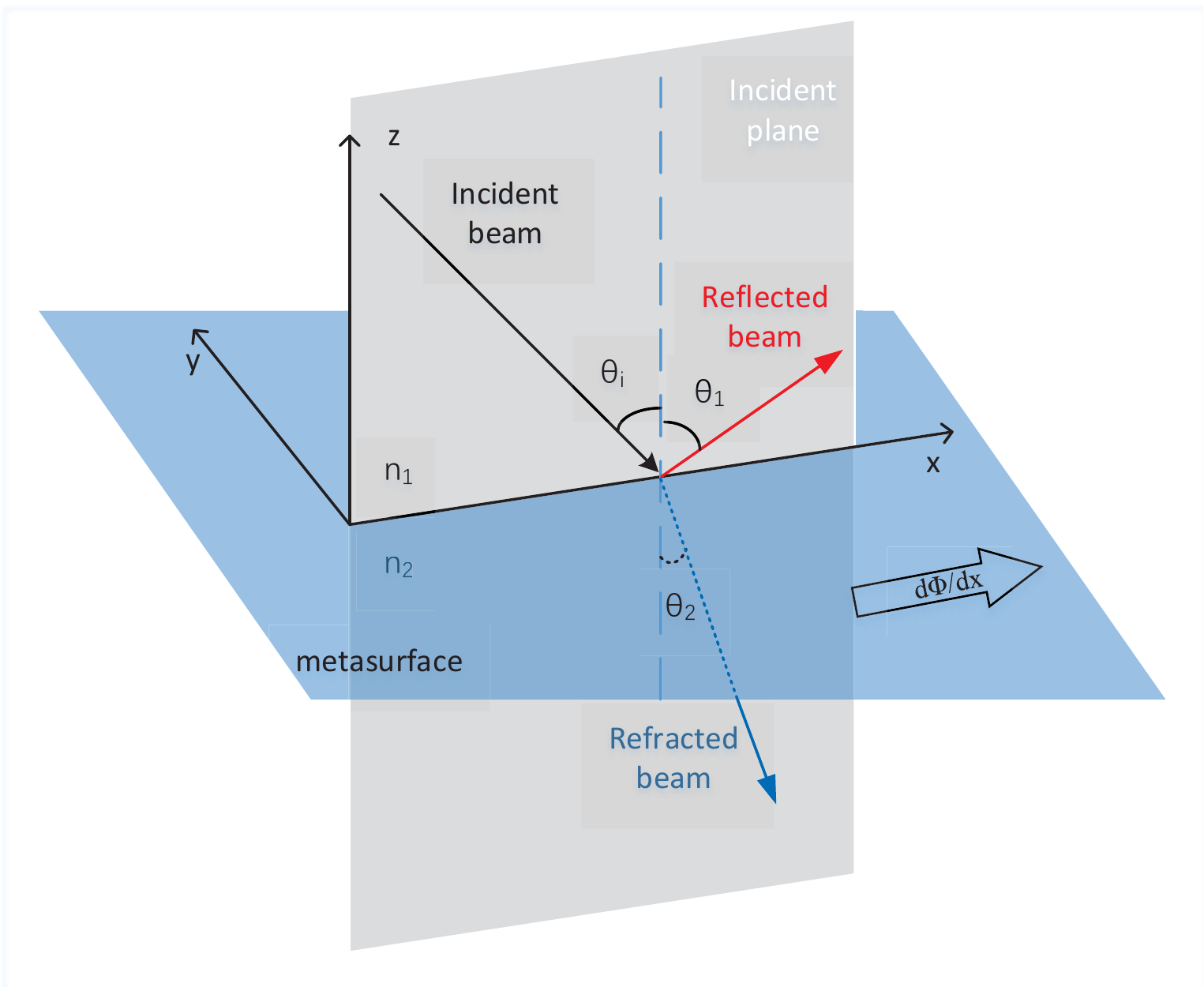}
\caption{Illustration of the generalized laws of refraction and reflection.}
\label{ref}
\end{figure}

\begin{pbox}
        \textit{Achieving anomalous reflection (Fig. \ref{func}(a))}\\

        Suppose that the phase discontinuity at the boundary is a function of the position along the $x$ direction $\Phi(\vec{r_x})$, where $\vec{r_x}$ is the position vector on the boundary. Moreover, suppose that the derivative of the phase discontinuity exists. Then, the angle of reflection ($\theta_1$) and the angle of refraction ($\theta_2$) are~\cite{chen2016review}:
        \begin{align}
        \label{gen_law}
            \theta_1 &= sin^{-1}\Big[sin \theta_i + \frac{\lambda}{2\pi n_1}\frac{d\Phi}{dx}\Big], \\
            \theta_2 &= sin^{-1}\Big[\frac{n_1}{n_2}sin \theta_i + \frac{\lambda}{2\pi n_2}\frac{d\Phi}{dx}\Big],
        \end{align}
        where $\theta_i$ is the angle of incidence, $\lambda$ is the wavelength of the transmitted signal in vacuum, and $n_1$, $n_2$ are the refractive indexes, as shown in Fig.~\ref{ref}.
\end{pbox}
There are other results related to the generalized laws of refraction and reflection, including the critical angles for total internal reflection or refraction. The main result just presented here states that, when a phase discontinuity is introduced at the boundary surface, the angles of reflection and refraction depend not only on the angle of incidence but also on the wavelength, refractive indexes, and the gradient of the phase discontinuity. This gives extra controllable parameters to manipulate the reflected and refracted EM waves. As a result, anomalous reflection can be achieved by tuning the phase gradient (${d\Phi}/{dx}$) based on \eqref{gen_law} or, in the discrete patch-array implementation, by tuning the length of the super-lattice. However, the assumption that the derivative of the phase discontinuity is constant ($d\Phi/dx = \text{const.}$) does not necessary hold if different wave transformations are needed.

\subsubsection{Co-phase condition}

Focusing is usually implemented when the RIS is within the near-field of the source or the terminal is close to the RIS. In these cases, the curvatures of the incident and reflected wavefront are non-negligible. The optimization of the surface aims to produce a pencil-beam pointing towards the direction of the terminal. When the link between the source and the RIS, as well as the link between the RIS and the terminal are in LoS, the following co-phase condition~\cite{huang2008reflectarray} can be applied.

\begin{figure}[t!]
\centering
\includegraphics[width =3.5in]{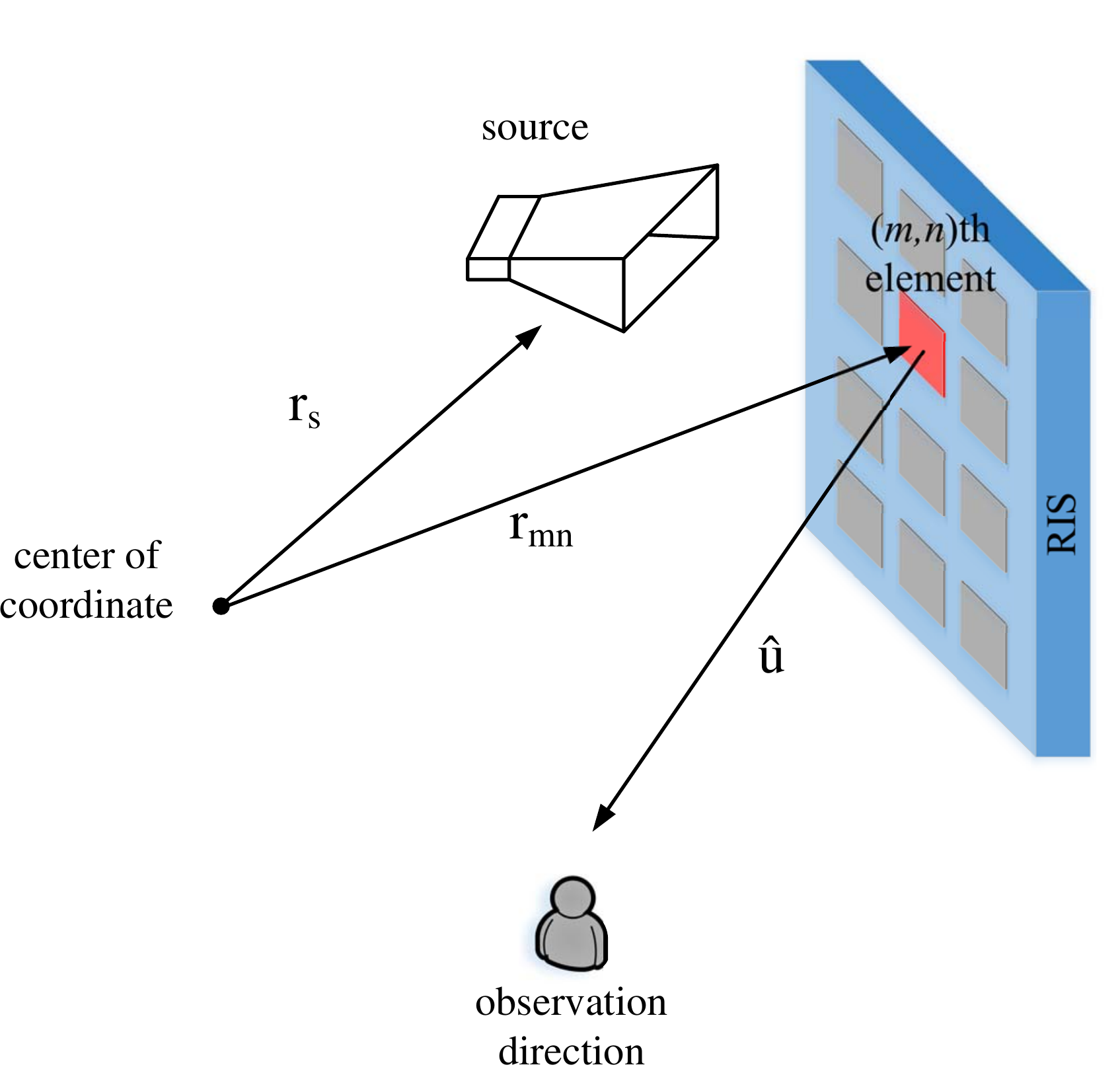}
\caption{Coordinate representation of co-phase condition.}
\label{coph}
\end{figure}

\begin{pbox}

        \textit{Achieving beamforming (focusing) (Fig. \ref{func}(b))}\\

Let $r_{mn}$ denote the position of the $(m,n)$-th RIS element, $r_s$ denote the position of the source, and $\hat{u}$ denote the direction of the observer with respect to the surface. As shown in Fig.~\ref{coph}, $\phi_{mn}$ can be chosen as follows~\cite{huang1998ka}:
\begin{equation}\label{cophase}
    -k_0(|\vec{r}_{mn}-\vec{r}_s|-\vec{r}_{mn}\cdot \hat{u})+\phi_{mn} = 2\pi \cdot t
\end{equation}
where $t=1,2,3...$ and $k_0 = 2\pi/\lambda_c$.
\end{pbox}

The two design principles just discussed provide guidance on how to configure the RIS phase shift patterns for typical applications. In more complex wireless communication systems, the RIS role is more intricate, thus cannot be categorized by the two working functions described in Fig.~\ref{func}. In these cases, to determine the RIS configuration, an optimization problem needs be formulated. These issues are elaborated in the following sections.

\subsection{Discussions and Outlook}

In the design and configuration of RISs, both theoretical limitations and hardware implementation limitations affect the overall performance of the resulting system. Theoretical limitations are, e.g., the result of considering simplified assumptions or adopting naive perspectives during the modeling of the RIS and its interaction with the EM wave. Hardware implementation limitations come, e.g., from the discretization of an ideally continuous RIS profile. In the following text, we elaborate on three major points.

\subsubsection{Hardware limitations}
In practical application scenarios, many hardware parameters significantly affect the achievable performance of the system. For example, the number of quantization levels of the RIS phase shifts, the maximum number of elements that is possible to integrate on the substrate, and the percentage of scattering environment that can be coated by an RIS. Existing research contributions studied the limitations and tradeoffs caused by these hardware constraints by analyzing their effect on the channel distribution~\cite{xu2020novel}, the power scaling law~\cite{qian2020beamforming}, and performance metrics such as the outage probability~\cite{zhang2019analysis} and ergodic capacity~\cite{boulogeorgos2020ergodic}. At the time of writing, the analysis of the impact of hardware limitations and how an RIS performs compared to other available technologies are open research issues.
\subsubsection{System design simplifications}
The adoption of oversimplified models for RIS hardware or channel models result in limitations on the system design. Because of the complex nature of RISs and their interaction with the environment, initial research contributions have adopted simple models. For example, using hardware models based on local designs, the ray-optics perspective for channel modeling, and decoupling reflection amplitude and phase shift of the RIS elements. Even though the ray-optics approximation can yield effective designs in some cases (as shown in Section \ref{prin}), it is preferable to adopt the wave-optics design in practical cases~\cite{Renzo_cylindrical_mirror_scatter}. Further research efforts are needed to bridge complex physical models of different RIS implementations with widely used communication models~\cite{gradoni2020end}.

\subsubsection{Optimization limitations}
To reap the benefits of deploying RISs in wireless networks, the RIS parameters (e.g., the reflection coefficient and deployment location) and network resource allocation (e.g., transmit beamforming and user scheduling) should be jointly optimized. However, the resulting joint optimization problems are normally non-convex and involve highly-coupled variables, which make it challenging to derive a globally optimal solution. Though some efficient algorithms have been recently proposed to compute high-quality suboptimal solutions~\cite{wu2019intelligent,huang2019reconfigurable,Mu_124}, the performance limits of RISs remain unknown. To fully understand the attainable performance limits, information-theoretic perspective investigations ~\cite{Karasik_Joint,Mu_Capacity} are important and sophisticated mathematical tools~\cite{liu2020ris} are expected to be employed. Further details are given in Sections VI and V.

\section{Performance Analysis of Multi-Antenna Assisted RIS Systems}\label{section:MIMO-RIS}

In Section II, we have discussed the fundamental physical properties of the RISs. However, how RISs affect the communication performance is still an open problem. To systematically survey existing designs for RIS-enhanced networks, we discuss the following topics: (1) channel models, (2) performance analysis, and (3) benchmark schemes.

\subsection{ \textbf{Channel Models}\label{large-scale-Hou}}

\textbf{1) Path Loss Models}

Some research contributions on the path loss for RISs are available in~\cite{di2020smart},~\cite{tang2019wireless} and~\cite{Renzo_cylindrical_mirror_scatter}, which showed that the power scattered by an RIS is usually formulated in terms of an integral that accounts, by leveraging the Huygens principle, for the impact of the entire surface in the free-space scenario, where scattering, shadowing, and reflection are ignored. Closed-form expressions of the integral are, on the other hand, difficult to obtain, except for some asymptotic regimes, which correspond to viewing the RIS as electrically small and electrically large (with respect to the wavelength and the transmission distances). It is worth mentioning, in addition, that the path-loss model depends on the particular phase gradient applied by the RIS. Notably, the scaling laws can be different if the RIS operates as an anomalous reflector and as a focusing lens. In the following, we briefly discuss two scaling laws that have recently been reported for RISs that operate as anomalous reflectors (described in the previous section). Further information and details can be found in~\cite{di2020smart} and~\cite{Renzo_cylindrical_mirror_scatter}.

\begin{itemize}
    \item \textbf{Electrically Small RISs:} In this asymptotic regime, the RIS is assumed to be relatively small in size compared with the transmission distances. In this regime, the RIS can be approximated as a small-size scatterer. In general, the path-loss scales with the reciprocal of the product of the distance between the transmitter and the center of the RIS and the distance between the center of the RIS and the receiver. In addition, the received power usually increases with the size of the RIS. The received power is usually maximized in the direction of anomalous reflection, where the path loss through the RIS follows the ``product of distances'' models, which can be formulated as:
        \begin{equation}\label{anomalous reflection}
        L({d_{SR}},{d_{RU}})\approx {\lambda_S}  ({d_{SR}}{d_{RU}})^{-1},
        \end{equation}
        where $\lambda_S$ denotes the coefficient of the electrically small scenario, $d_{SR}$ and $d_{RU}$ represent the distance of source-RIS and RIS-user links, respectively. A detailed discussion is given in~\cite[Secs. IV-A, IV-B, IV-C]{Renzo_cylindrical_mirror_scatter}.
    \item \textbf{Electrically Large RISs:} In this asymptotic regime, the RIS is assumed to be large (ideally infinitely large) in size compared with the transmission distances and the wavelength. In this regime, the RIS can be approximated as a large flat mirror. Let us denote by $x_0$ the point of the RIS (if it exists) at which the first-order derivative of the total phase response of the combined incident signal, reflected signal, and the surface reflection coefficient of the RIS is equal to zero. In general, the path-loss asymptotically scales with the reciprocal of a weighted sum of the distance between the transmitter and $x_0$ and the distance between $x_0$ and the receiver. In addition, the received power is not dependent on the size of the RIS, which is viewed as asymptotically infinite. This result substantiates the fact that the power scaling law of the RIS is physically correct, since it does not grow to infinity as the size of the RIS goes to infinity. This is because the scaling law and the behavior of the RIS are different with respect to the electrically small regime. In this case, the path loss through the RIS follows the ``sum of distances'' models, which can be approximated as:
        \begin{equation}\label{Elec_large}
        L({d_{SR}},{d_{RU}})\approx {\lambda_L} (d_{SR}+d_{RU})^{-1},
        \end{equation}
        where $\lambda_L$ denotes the coefficient of the electrically large scenario. A detailed discussion is given in~\cite[Secs. IV-A, IV-B, IV-C]{Renzo_cylindrical_mirror_scatter}.
\end{itemize}

\textbf{2) Spatial Models}

Stochastic geometry (SG) tools are capable of capturing the location randomness of users thus enabling the derivation of computable or closed-form expressions of key performance metrics. Specifically, several spatial processes exist for modeling the locations of users in different wireless networks, i.e., the homogeneous Poisson point process (HPPP)~\cite{Liu_Coop_NOMA_SWIPT,Liu_physical_scurity_NOMA}, the Poisson cluster process (PCP)~\cite{PCP_Yi,PCP_Hou}, the Binomial point process (BPP), as well as the Hard core point process (HCPP)~\cite{SG_survey}. We list some promising approaches for analyzing the performance of RIS-enhanced networks by using SG in Table~\ref{table:RIS_SG}.
In~\cite{stochastic_twoLayer_PPP}, the RIS elements are employed on obstacles, and it is assumed that randomly distributed users are located in the serving area of the RISs. In~\cite{Renzo_random_RIS}, the objects are modeled by a modified random line process of fixed length and with random orientations and locations. Therein, the probability that a randomly distributed object that is coated with an RIS acts as a reflector for a given pair of transmitter and receiver was investigated. In~\cite{ChaoZ2021_PCP_RIS}, PCP was invoked in the RIS-enhanced large-scale networks, where the angle of reflection is constrained by the angle of incidence. Therefore, the randomly distributed users and BSs are located at the same side of the RIS.

\begin{table*}[htbp]\scriptsize
\caption{Summary of RIS-enhanced SG Networks}
\begin{center}
\centering
\begin{tabular}{|l|l|l|l|}
\hline
\centering
\textbf{Approaches} & \textbf{Advantages} & \textbf{Disadvantages} &  \textbf{Ref.}  \\
\hline
\centering
RIS-enhanced HPPP & Fairness-oriented design & Restricted user distribution & \cite{hou2019mimo}\\
\hline
\centering
RIS-enhanced two layer HPPP &  Coverage-hole enhancement & RISs are deployed at obstacles  & \cite{stochastic_twoLayer_PPP}\\
\hline
\centering
HPPP conditioned on angle & Practical design & Not tractable & \cite{NOMA_poission}\\
\hline
\centering
RIS-enhanced HPPP conditioned on angle & Practical design for RIS-enhanced networks & Complicated & --\\
\hline
\end{tabular}
\end{center}
\label{table:RIS_SG}
\end{table*}

\textbf{3) Small-Scale Fading Models}

Currently, two main approaches have been used for analyzing the performance of RIS-aided systems in small-scale fading channels: (1) the central-limit-theorem-based (CLT-based) distribution and (2) the use of approximated distributions.
\begin{itemize}
    \item \textbf{CLT-based Distribution:} Let us consider a single-antenna BS that communicates with a single-antenna user with the aid of an RIS of $N$ elements. If the two received signals, from the BS and from the RIS, can be coherently combined, the effective channel power gain is given by
        \begin{equation}\label{Small_scale_CLT}
        \begin{aligned}
         & {\left| {{{\mathbf{r}}^H}{\mathbf{\Phi g}} + h} \right|^2}, \\
         {\rm{s.t.}}\;\;& {\beta _{1}}, \cdots, {\beta _{N}} = 1\\
         & {\phi _{1}}, \cdots, {\phi _{N}} \in \left[ {0,2\pi } \right),
         \end{aligned}
        \end{equation}
        where $h\in {{\mathbb{C}}^{1 \times 1}}$, ${\mathbf{g}} \in {{\mathbb{C}}^{M \times 1}}$, ${\mathbf{r}} \in {{\mathbb{C}}^{M \times 1}}$ denote the channels of the BS-user, BS-RIS, and RIS-user links, respectively. ${\mathbf{\Phi }} = {\rm{diag}}\left( {{\beta _1}{e^{j{\phi _1}}},{\beta _2}{e^{j{\phi _2}}}, \ldots ,{\beta _N}{e^{j{\phi _N}}}} \right)$ denotes the reflection-coefficient matrix of the RIS, where $\left\{ {{\beta _1},{\beta _2} \ldots ,{\beta _N}} \right\}$ and $\left\{ {{\phi _1},{\phi _2} \ldots ,{\phi _N}} \right\}$ represent the amplitude coefficients and phase shifts of the RIS elements, respectively. In this setup, the CLT-based technique stands as an approximation tool for analyzing the performance in the low-medium-SNR regime. This is due to the fact that the distribution of the probability density function (PDF) in the range $0$ to $0+$ is not precise by using the CLT~\cite{DING_SISO_CLT}.
        In Rayleigh fading channels, the distribution of an RIS-enhanced link follows a modified Bessel function~\cite{DING_SISO_CLT}. Since both transmitter and RISs are part of the infrastructure, and the RISs are typically positioned to exploit the LoS path with respect to the locations of the transmitters and the receivers for increasing the received signal power, Zhang~\emph{et al.}~\cite{zhang2019analysis} studied Rician fading channels, and the analysis showed that the signal power follows a non-central chi-squared distribution with two degrees of freedom. Ding and Poor~\cite{RIS_zhiguo_simple} proposed an RIS-enhanced network, where RISs are utilized for effectively aligning the directions of the users' channel gains. By utilizing the CLT-based technique, Cheng~\emph{et al.}~\cite{Yanyu2021_multi_RIS} studied the multi-RIS network, where the channel distributions were investigated with or without BS-user links.
    \item \textbf{Approximated Distribution:} The exact distribution of the received SNR of the signal reflected from an RIS is non-trivial to be obtained, and hence the use of approximated distributions is often necessary.
Qian~\emph{et al.}~\cite{qian2020beamforming} proposed a simple approximated distribution of the received SNR, and proved that the received SNR can be approximated by two (or one) Gamma random variables and the sum of two scaled non-central chi-square random variables. A prioritized signal enhancement design was proposed by Hou~\emph{et al.}~\cite{Hou_SISO_NOMA_RIS}, where both the outage performance and ergodic rate of the user with the best channel gain were calculated. Lyu and Zhang~\cite{lyu2020spatial} proposed a single-input and single-output (SISO) network with multiple randomly deployed RISs, and showed that the exact distribution in terms of received signal power can be approximated by a Gamma distribution. Makarfi~\emph{et al.}~\cite{RIS_fading_to_gamma} proposed an RIS-enhanced network, whose equivalent channel is modeled by the Fisher-Snedecor $\mathcal{F}$ distribution.
\end{itemize}

Based on the above mentioned contributions~\cite{DING_SISO_CLT,zhang2019analysis,RIS_zhiguo_simple,Hou_SISO_NOMA_RIS,lyu2020spatial,RIS_fading_to_gamma}, where only approximated channel distributions are obtained, the exact channel distribution of RIS-enhanced networks is still an open problem. Based on recent research results, for example, a fundamental limitation lies in the calculation of the diversity order of RIS-enhanced networks under ideal operating conditions and in the presence of hardware limitations, e.g., quantized phase shifts. For example, the diversity order obtained by using the CLT-based distribution~\cite{DING_SISO_CLT} is $\frac{1}{2}$ in the high-SNR regime, whereas the diversity order is $\frac{N}{3}$ if an approximation based on the Gamma distribution is used, where $N$ denotes the number of RIS elements. However, the CLT-based and Gamma-based distributions are not exact, making the performance analysis of RIS-enhanced networks an interesting problem for future research.
Furthermore, since the exact distribution contains higher-order components, which approach zero in the high-SNR regime, most of the previous contributions~\cite{cheng2020downlink,Tang2020PLS} adopt the approximated distribution method for modeling small-scale fading channels, and the exact distribution of RIS-enhanced networks is still an open problem. For example, recent exact results on the impact of phase noise on the diversity order of RIS-enhanced transmission can be found in~\cite{discrete_RIS_order}.

\subsection{Performance Analysis}

In this subsection, we briefly discuss currently available papers on the performance analysis of RISs that are realized as large arrays of tiny and inexpensive antennas whose phase response is locally optimized. By offering extra diversity in the spatial domain, multi-antenna techniques are of significant importance. The application of multi-antenna enhanced RIS networks has attracted substantial interest from academia~\cite{lyu2020spatial,Hou_SISO_NOMA_RIS,hou2019mimo} and industry~\cite{RIS_industrial_linglong,Hou_MIMO_RIS_IC,Bjrnson2019}.
Given the increasing number of research contributions on RISs, its advantages are becoming more clear, especially in terms of spectral efficiency (SE) and energy efficiency (EE) enhancement. There are several key challenges for performance analysis in RIS-enhanced networks. One of the main challenges is to evaluate the exact distributions of the cascade channels between the BS and users through RISs. Another challenge is evaluating the effective channel gain after passive beamforming at the RIS. Table~\ref{table:RIS} summarizes the existing contributions on RISs with multiple antennas and illustrates their comparison. RIS-enhanced single user networks have been analyzed in \textbf{Section~III.A}. Hence, we turn out attention to RIS-enhanced multi-user networks. A prioritized signal-enhancement-based (SEB) was proposed by Hou~\emph{et al.}~\cite{Hou_SISO_NOMA_RIS}, where passive beamforming is designed for the user with the best channel gain, and all the other users rely on RIS-enhanced beamforming.

\begin{table*}[htbp]\scriptsize
\caption{Important contributions on RIS-enhanced networks. ``DL'' and ``UL'' represent downlink and uplink, respectively. The ``sum-rate gain'' implies that the gain brought by invoking RIS technique}
\begin{center}
\centering
\begin{tabular}{|l|l|l|l|l|l|}
\hline
\centering
\textbf{Ref.} &\textbf{Scenarios} & \textbf{Direction}& \textbf{Users} & \textbf{Main Objectives} & \textbf{Techniques} \\
\hline
\centering
\cite{hou2019mimo} & MIMO & DL & Multiple users & OP and throughput & Fairness SEB \\
\hline
\centering
\cite{Bjrnson2019} & SISO & DL & Single user & sum-rate gain & Compare with relay\\
\hline
\centering
\cite{RIS_fading_to_gamma} & SISO & UL & Single user & OP and throughput & Effective channel gain \\
\hline
\centering
\cite{DING_SISO_CLT} &  SISO & DL & Single user & Effective channel gain & Compare with random phase shifting\\
\hline
\centering
\cite{zhang2019analysis} &  SISO & DL & Single user & Effective channel gain & SEB\\
\hline
\centering
\cite{RIS_zhiguo_simple} & SISO & DL & Single user & OP & SEB\\
\hline
\centering
\cite{Hou_SISO_NOMA_RIS} & SISO & DL & Multiple users &  OP and throughput & Prioritized SEB \\
\hline
\centering
\cite{lyu2020spatial} &  SISO & DL & Single user  & Effective channel gain & SEB\\
\hline
\centering
\cite{qian2020beamforming} & MIMO & DL & Single users  & Effective channel gain  & Random matrix theory and CLT \\
\hline
\centering
\cite{Ntontin2019multi}& MIMO & DL & Multiple users  & sum-rate gain  & SEB  \\
\hline
\centering
\cite{Hou_MIMO_RIS_IC} & MIMO & DL  & Multiple users & Interference cancellation & SCB and less constraint at RAs\\
\hline
\centering
\cite{discrete_RIS_number_bit} & SISO & UL & Multiple users & Sum-rate & Minimum required finite resolution\\
\hline
\centering
\cite{CR_RIS_Rice} & MISO & DL & Multiple users & Sum-rate & Multi-RIS distribution\\
\hline
\centering
\cite{RIS_ruiZhang_discrete} & MISO & DL & Multiple users & Sum-rate & Discrete phase shifts \\
\hline
\end{tabular}
\end{center}
\label{table:RIS}
\end{table*}

\begin{itemize}
        \item \textbf{RIS-enhanced Signal Enhancement Designs:} By assuming that multiple waves are co-phased at the users, the received signal can be significantly enhanced, which leads to the following optimization problem:
            \begin{equation}\label{SEB}
            \begin{aligned}
             {\rm{max}}\;\;&{\left| {{{\mathbf{r}}^H}{\mathbf{\Phi g}} + h} \right|^2} \\
           {\rm{s.t.}}\;\; & {\beta _{1}}, \cdots, {\beta _{N}} = 1\\
            & {\phi _{1}}, \cdots, {\phi _{N}} \in \left[ {0,2\pi } \right).
            \end{aligned}
            \end{equation}
            In order to further enhance the SE of RIS-enhanced networks, multiple antenna techniques can be employed at both the BS and users. Yuan~\emph{et al.}~\cite{CR_RIS_Rice} proposed a cognitive-radio-based RIS-enhanced multiple-input and single-output (MISO) network, where both perfect and imperfect channel state information (CSI) setups were considered. However, in many research works, continuous amplitude coefficients and phase shifts are assumed at the RISs, whilst in practice the phase shifts of RISs may not be continuous. Thus, You~\emph{et al.}~\cite{RIS_ruiZhang_discrete} proposed a discrete phase shifts model for a MISO enhanced RIS network. Zhang~\emph{et al.}~\cite{discrete_RIS_number_bit} then evaluated the required number of bits for finite-resolution RISs in an uplink SISO network. Hou~\emph{et al.}~\cite{hou2019mimo} investigated an RIS-enhanced MIMO network, where a fairness oriented design was considered by applying SG tools for modeling the impact of the users' locations.
        \item  \textbf{RIS-enhanced Signal Cancellation Designs:} Another application of deploying RISs in wireless networks is signal cancellation, where the reflected signals and the direct signals can be destructively combined. The corresponding optimization problem can be formulated as follows:
            \begin{equation}\label{SCB}
            \begin{aligned}
            {\rm{min}}\;\;&{\left| {{{\mathbf{r}}^H}{\mathbf{\Phi g}} + h_I} \right|^2} \\
            {\rm{s.t.}}\;\; & {\beta _{1}}, \cdots, {\beta _{N}}  \le  1\\
            & {\phi _{1}}, \cdots, {\phi _{N}} \in \left[ {0,2\pi } \right).
            \end{aligned}
            \end{equation}
            where $h_I$ denotes the aggregate interference signals from other-cell BSs. By assuming that both the inter-cell and intra-cell interferences are perfectly known, the optimal solution to~\eqref{SCB} is to adjust both the signal phase and amplitude coefficients of the BS-RIS-user links to the opposite of the effective interference links with the same amplitude. By doing so, some promising applications can be realized, e.g. RIS-enhanced PLS and interference cancellation. On the one hand, by assuming that perfect CSI is available at the RIS controller, the inter-cell and intra-cell interferences can be eliminated. On the other hand, considering the PLS requirements, RISs also stand as a potential solution for cooperative jamming techniques, i.e., RISs act as artificial noise sources. By adopting this approach, several contributions have been made. Hou~\emph{et al.}~\cite{Hou_MIMO_RIS_IC} proposed an RIS-enhanced interference cancellation technique in a MIMO network, where the inter-cluster interference can be eliminated without active beamforming weights and detection vectors. Furthermore, this work can be adopted for application to coordinated multi-point (CoMP) networks for inter-cell interference cancellation in cellular networks. Shi~\emph{et al.}~\cite{PLS_RIS_SWIPT} investigated an RIS-enhanced secure beamforming technique, where the secrecy rate of the legitimate user was derived. Lyu~\emph{et al.}~\cite{PLS_RIS_Jammer} investigated an RIS jamming scenario, where RISs act as jammers for attacking legitimate communications without using any internal energy.
\end{itemize}

\subsection{Benchmark Schemes}

\begin{figure}[t!]
\centering
\includegraphics[width =3in]{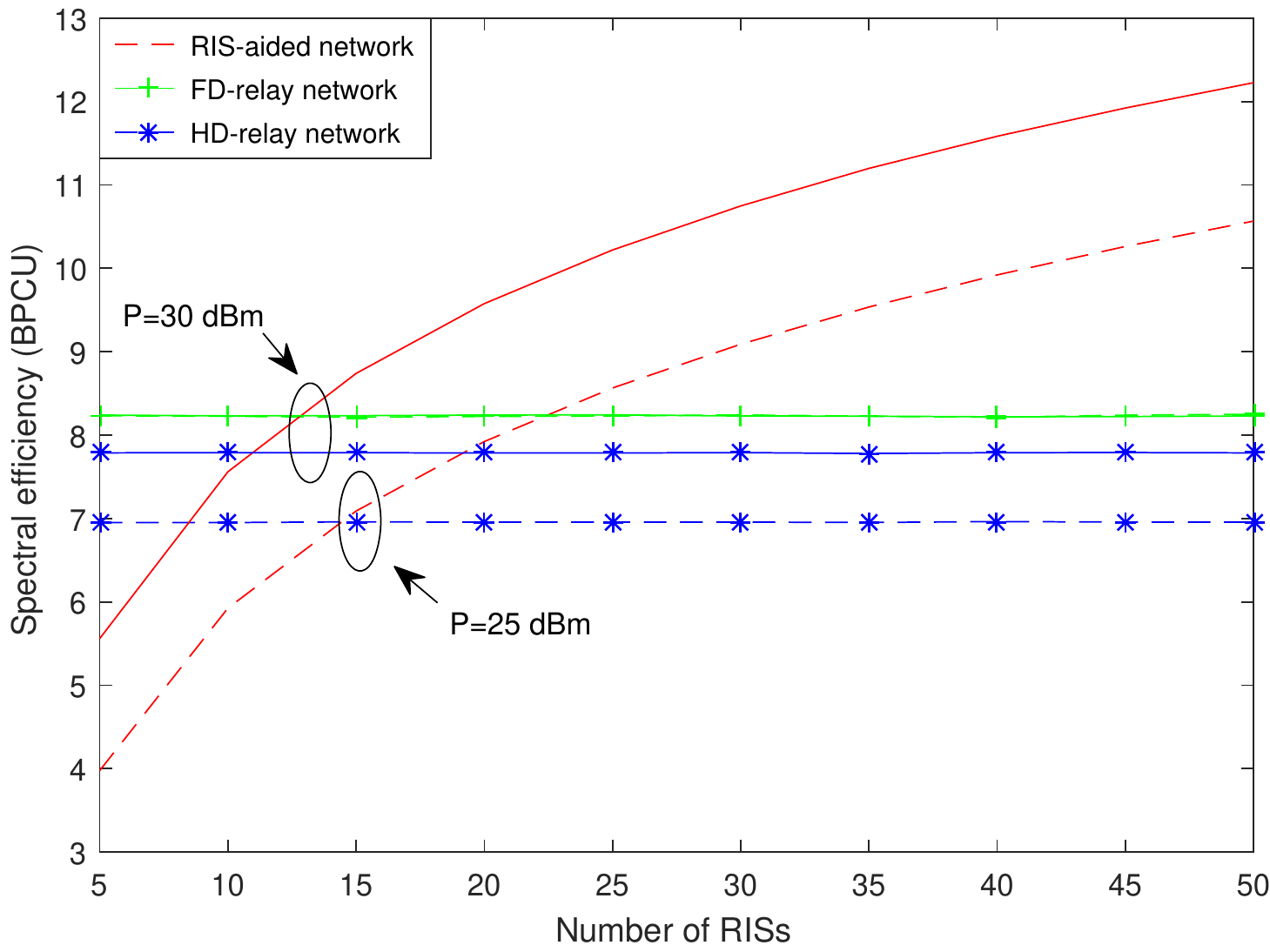}
\caption{Spectral efficiency of RIS-enhanced, FD-relay as well as HD-relay networks versus the number of RIS elements. Please refer to~\cite{Hou_SISO_NOMA_RIS} for simulation parameters.}
\label{Compare With AF_DF relay fading fig 6}
\end{figure}

In order to assess the advantages and limitations of RISs, two benchmark transmission technologies are usually considered: 1) surfaces with random phase shifts; and 2) relay networks.
\begin{itemize}
        \item \textbf{Random Phase Shifts:} RISs are capable of shifting the phase of the incident signal, and hence multiple signals can be boosted or eliminated at the user side or at the BS side. Hence, a well-accepted benchmark scheme to quantify the performance enhancement by RIS elements is given by a surface that is not configurable and that can ideally be modeled by a surface with random phase shifts~\cite{DING_SISO_CLT}.
        \item \textbf{Relay Networks:} Generally speaking, relay-aided networks can be classified into two pairs of classic relaying protocols, which are 1) FD and HD relay networks; and 2) AF and DF relay networks. By assuming that the optimal power split strategies of both the AF and DF relays are employed, the performance gain between RIS-enhanced and relay-aided networks can be compared. Specifically, Bjornson~\emph{et al.}~\cite{Bjrnson2019} compared the achievable data rate of both RIS-enhanced and DF-relay-aided SISO network, where the BS-user links are blocked. It was pointed out that when the number of tunable elements of the RISs is large enough, an RIS-enhanced network is capable of outperforming a DF-relay-aided network. In an effort to provide a comprehensive analysis for both RIS-enhanced and relay-aided networks, Ntontin~\emph{et al.}~\cite{Ntontin2019multi} compared the system performance of classic maximal ratio transmission (MRT) and maximal ratio combining (MRC) techniques. Fig.~\ref{Compare With AF_DF relay fading fig 6} illustrates the potential benefits of RIS-enhanced networks compared with both HD-relay and FD-relay networks in terms of network throughput~\cite{Hou_SISO_NOMA_RIS}. Here, the performance of HD-relaying is obtained for an equal time-split ratio. We can see that the network throughput gap between the RIS-enhanced network and the other pair of relay aided networks becomes smaller, when the number of RIS elements increases. For example, when the number of RIS elements $N=23$ and the transmit power $P=25$ dBm, the proposed RIS-enhanced network is capable of outperforming both FD and HD relay aided networks, which indicates that the RIS-enhanced network becomes more competitive, when the number of RIS elements is large enough.
            Table~\ref{table:RIS_compare_benchmarks} provides a comparison between RIS-enhanced and relay aided networks in terms of advantages and limitations.
\end{itemize}

\begin{table*}[htbp]\scriptsize
\caption{Comparison of the RIS-enhanced and Benchmarks. The ``Power'' Implies Additional Power Supply at the RISs or at the Relay. The ``CT'' Denotes Concurrent Transmission}
\begin{center}
\centering
\begin{tabular}{|l|l|l|c|c|c|c|}
\hline
\centering
\textbf{Mode} & \textbf{Pros} & \textbf{Cons} &\textbf{Delay}&  \textbf{Power} & \textbf{Interference} & \textbf{CT}   \\
\hline
\centering
RIS-enhanced  & High EE, simple device & CSI must be perfectly known & \xmark & \xmark & \xmark & \checkmark  \\
\hline
\centering
HD relaying under AF protocol & No decoding at relay & Noise is amplified  & \checkmark & \checkmark & \xmark & \xmark \\
\hline
\centering
HD relaying under DF protocol & No self-interference & Latency is high & \checkmark & \checkmark & \xmark & \xmark \\
\hline
\centering
FD relaying under AF protocol & No decoding at relay & Noise and interference are amplified & \xmark & \checkmark & Self-interference & \checkmark \\
\hline
\centering
FD relaying under DF protocol & No latency & Rate ceiling occurs & \xmark & \checkmark & Self-interference & \checkmark \\
\hline
\centering
MIMO relay & High SE & High cost, difficult to realize at mmWave & \xmark & \checkmark & \checkmark & \checkmark \\
\hline
\end{tabular}
\end{center}
\label{table:RIS_compare_benchmarks}
\end{table*}

\subsection{Discussions and Outlook}
Although previous research contributions have analyzed the approximated performance of RIS-enhanced networks, there are still three major open research problems.

1) Path loss experiments for outdoor scenarios: Since only~\cite{tang2019wireless} reported experimental measurements of the path loss in free-space environments, the development of experimentally-validated path-loss models in outdoor scenarios is an open research issue, especially in the presence of reflecting and scattering objects.

2) Exact distributions: Current research methods for performance analysis are based on the CLT-based distribution and approximated distribution~\cite{Yanyu2021_multi_RIS,RIS_fading_to_gamma}, which are, however, accurate only in the high-SNR regime. More advanced and accurate analytical models are needed for analyzing the diversity order and the network performance in the low-SNR regime.

3) Integrated application scenarios: To serve the desired users in different scenarios, signal enhancement and signal cancellation designs were well investigated in recent research works~\cite{Hou_SISO_NOMA_RIS,Ntontin2019multi,Hou_MIMO_RIS_IC}. However, the desired signals and interference signals can be simultaneously enhanced and mitigated, which constitutes an important future direction.

\section{RIS Beamforming and Resource Allocation}
As described in the previous section, deploying RISs enables high performance enhancements in wireless networks. Motivated by this benefit, how to jointly design the transmit and passive beamforming as well as the optimal allocation of the wireless resources has become an important task for RIS-enhanced wireless networks. In this section, we first discuss the information-theoretic performance limits of RISs. Then, we review recent research contributions with a particular focus on the joint beamforming optimization and the resource allocation design. Along the literature review, representative mathematical tools for facilitating the two types of design are also discussed along with their benefits and drawbacks.
\subsection{Information-Theoretic Perspective}
In order to unveil the fundamental performance limits of RISs, several research works~\cite{Karasik_Joint,Mu_Capacity,Zhang_MAC} have been devoted to investigating the RIS performance gains from an information-theoretic perspective.
\begin{figure}[t!]
\centering
\subfigure[Capacity regions with RIS-NOMA.]{\label{capacityNOMA}
\includegraphics[width= 3in]{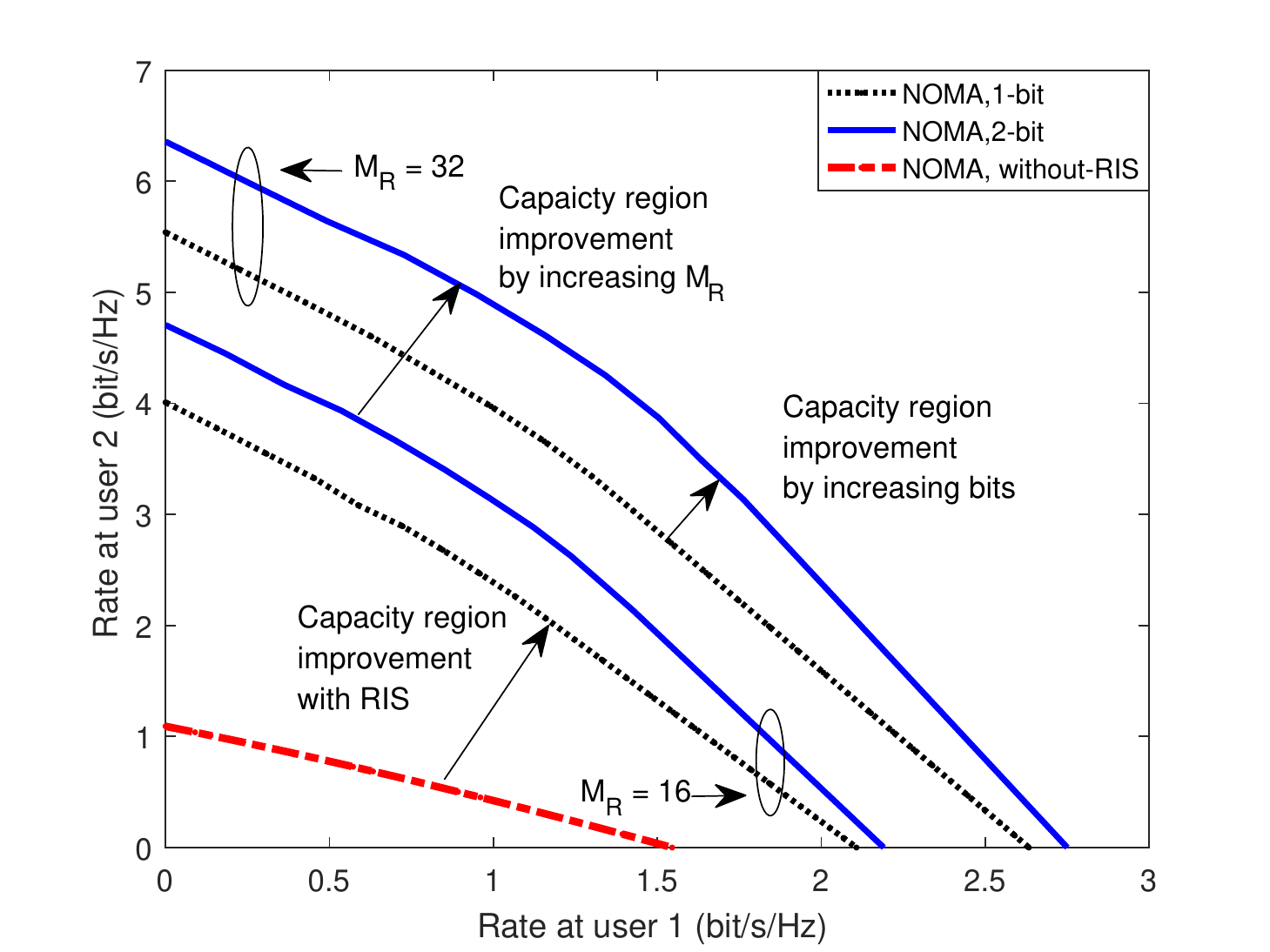}}
\subfigure[Rate regions with RIS-OMA.]{\label{rateOMA}
\includegraphics[width= 3in]{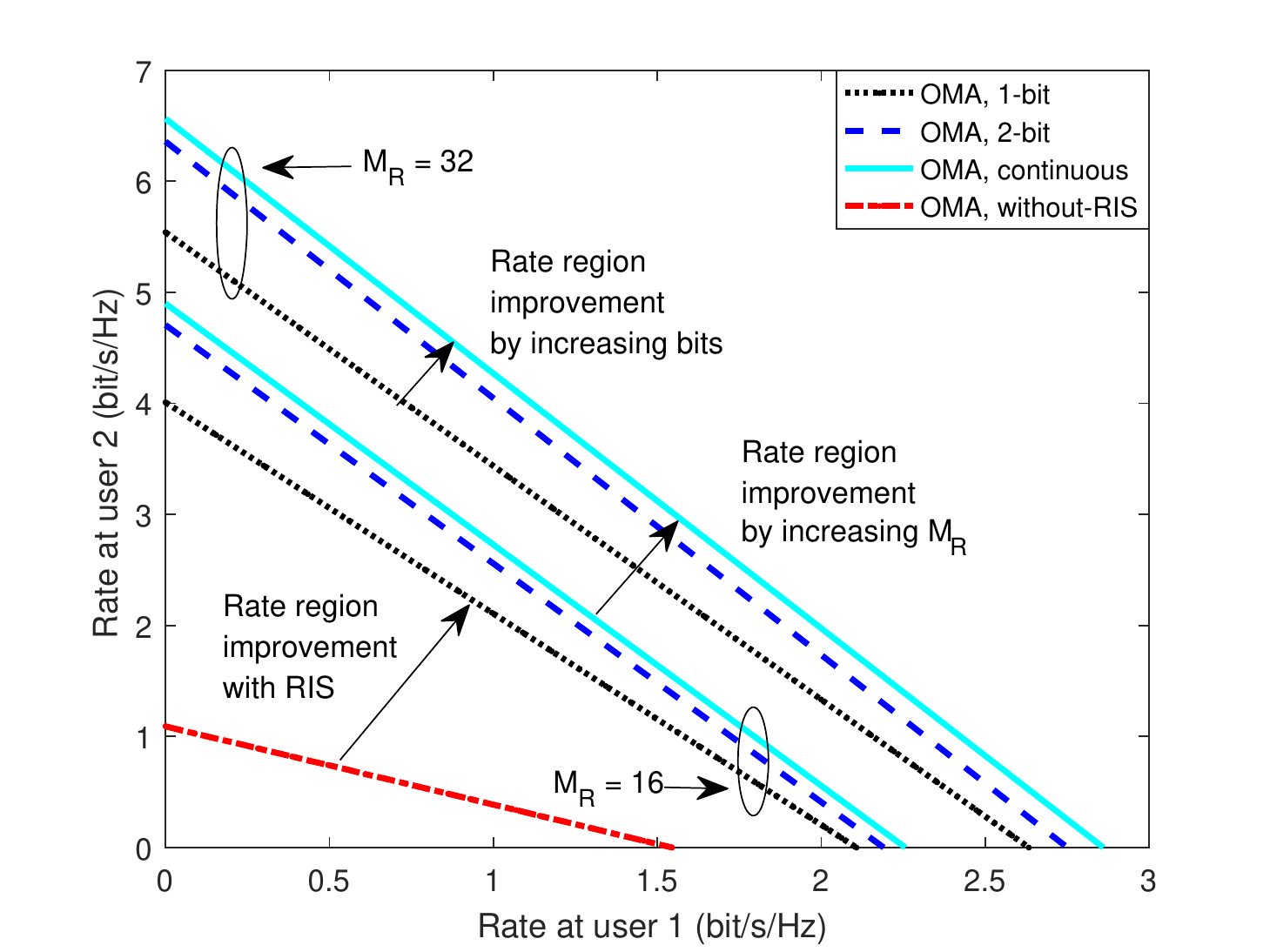}}
\setlength{\abovecaptionskip}{-0cm}
\caption{Illustration of the capacity and rate regions for a random channel realization with different RIS phase resolutions. $M_R$ denotes the number of RIS reflection elements. The full parameter settings can be found in \cite{Mu_Capacity}.}\label{RIS capacity}
\end{figure}
\begin{itemize}
  \item \emph{Capacity-achieving design:} In~\cite{Karasik_Joint}, Karasik {\em et al.} derived the capacity for an RIS-aided single-input and multiple-output (SIMO) communication system. With finite input signal constellations, it was proved that a joint information encoding scheme at both the transmitted signals and the RIS configurations is necessary for achieving the channel capacity~\cite{Karasik_Joint}. Based on this insight, the authors further proposed a practical transmission strategy by utilizing layered encoding and successive cancellation decoding techniques. Numerical examples showed that the proposed joint encoding scheme outperforms the conventional max-SNR scheme.
  \item \emph{Capacity region characterization:} The capacity region of the fading SISO broadcast channel (BC) was proved to be achieved by invoking the superposition coding (SC) at the transmitter and the successive interference cancellation (SIC) at multiple receivers~\cite{Ergodic}, i.e., employing the NOMA transmission. Inspired by these results, Mu {\em et al.}~\cite{Mu_Capacity} investigated the capacity and rate regions of RIS-enhanced multi-user wireless communication systems achieved by NOMA and orthogonal multiple access (OMA), respectively. The Pareto boundary of each region was characterized by solving a series of sum rate maximization problems via the rate-profile technique. As shown in Fig. \ref{capacityNOMA} and Fig. \ref{rateOMA}, by deploying an RIS, the NOMA capacity region and the OMA rate region can be improved. The capacity/rate regions are further enlarged by employing more precise phase resolutions and larger numbers of reflection elements. Furthermore, Zhang {\em et al.}~\cite{Zhang_MAC} investigated the capacity region of the multiple access channel (MAC) with two users in both centralized and distributed RIS deployment strategies. The results demonstrated that the centralized RIS deployment strategy can achieve a higher capacity gain than the distributed strategy.
\end{itemize}

\indent
\subsection{Joint Transmit and Passive Beamforming Design}
\begin{figure}[t!]
    \begin{center}
        \includegraphics[width=2.8in]{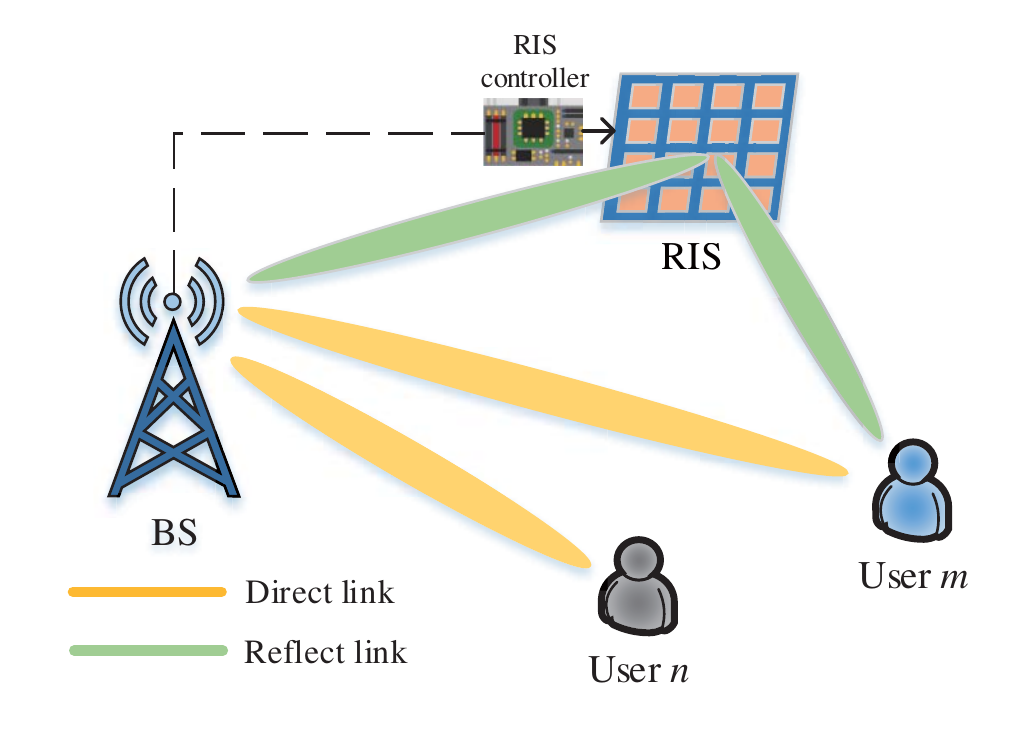}
        \setlength{\abovecaptionskip}{-0cm}
        \caption{Illustration of joint transmit and passive beamforming design.}
        \label{BF}
    \end{center}
\end{figure}
\subsubsection{Optimization objectives}As shown in Fig. \ref{BF}, an RIS is deployed to assist the transmission between the BS and the users by passively reflecting the signals. The RIS reflection coefficients can be adjusted by the BS through an RIS controller. Hence, the transmit beamforming at the BS and the passive beamforming at the RIS have to be jointly designed to improve the communication performance. In the following, we review the related research works in terms of their considered optimization objectives.
\begin{itemize}
\item \emph{Transmit power minimization or EE maximization:} In~\cite{wu2019intelligent}, Wu {\em et al.} minimized the transmit power for an RIS-enhanced MISO system in both single-user and multi-user scenarios. Alternating optimization (AO) based algorithms were developed to find locally-optimal solutions. The passive beamforming was designed by invoking the semidefinite relaxation (SDR) approach. It was revealed that an RIS can simultaneously enhance the desired signal strength and mitigate the interference for the multi-user scenario. The same problem was further investigated in~\cite{Wu_138} by taking discrete RIS phase shifts into consideration. The optimal solutions were derived by applying the branch-and-bound method and exhaustive search for single-user and multi-user scenarios, respectively. To reduce the computational complexity, efficient successive refinement algorithms were further designed. It was shown that the proposed low complexity algorithms are capable of achieving near-optimal performances. Han {\em et al.}~\cite{han2019intelligent} investigated physical-layer broadcasting in an RIS-aided network, where the total transmit power for satisfying QoS requirements of all users was minimized. Furthermore, Fu {\em et al.}~\cite{Fu_142} focused on an RIS-enhanced MISO downlink NOMA system, where the transmit power was minimized by jointly optimizing the transmit and passive beamforming vectors as well as user decoding orders. In order to overcome the drawbacks of the SDR approach, an alternating difference-of-convex (DC) method was proposed for handling the non-convex rank-one constraint. Zhu {\em et al.}~\cite{Zhu_NOMA} proposed an improved quasi-degradation condition for the RIS-enhanced MISO NOMA system to minimize the transmit power. Under this condition, NOMA is shown to be able to outperform the zero-forcing beamforming scheme. Zheng {\em et al.}~\cite{Zheng_NOMA} compared the minimum transmit power performance between OMA and NOMA in a discrete phase shift RIS-enhanced SISO system. A near-optimal solution was obtained by applying the linear approximation initialization and the AO method. The results showed that NOMA may perform worse than TDMA when the users have symmetric deployments and rate requirements, which revealed the importance of user pairing in the RIS-assisted NOMA system. Huang {\em et al.}~\cite{huang2019reconfigurable} solved the EE maximization problem in an RIS-enhanced multi-user MISO system, where a realistic RIS power consumption model was proposed in terms of the number of reflection elements and the phase resolutions at the RIS. An AO-based algorithm was designed for addressing the formulated problem, where the RIS phase shifts and the BS transmit power allocation were optimized by invoking the gradient descent method and the fractional programming method. The results demonstrated that the RIS achieves significantly better EE performance than the traditional active relay-assisted communication. In contrast to the aforementioned works based on the perfect CSI assumption, Zhou {\em et al.}~\cite{Zhou_141} investigated the robust beamforming design for an RIS-enhanced multi-user MISO system with imperfect CSI assumptions. The transmit power was minimized while satisfying QoS requirements of all users under all possible channel error realizations. The formulated non-convex problem was transformed into a sequence of semidefinite programming (SDP) subproblems, where the CSI uncertainties and the non-convex unit-modulus constraints were handled by applying approximation transformations and the convex-concave procedure~\cite{lipp2016variations}, respectively. In~\cite{Zhou_IP}, the robust beamforming design was further studied under two channel error models, namely the bounded CSI error model and the statistical CSI error model. The S-procedure and the Bernstein-Type inequality were applied in each model. The results unveiled that the RIS may degrade the system performance when the channel error is high. Zappone {\em et al.}~\cite{zappone2020overhead} modeled the overhead for carrying out the channel estimation and adjusting the RIS. Based on the proposed overhead model, the EE of an RIS-empowered MIMO communication network was maximized by jointly optimizing the RIS phase shifts as well as the transmitted and received filters.

\item \emph{SE or capacity maximization:} Yu {\em et al.}~\cite{Yu140} considered the SE maximization problem in an RIS-enhanced MISO system. Since the SDR approach only provides an approximate solution~\cite{wu2019intelligent}, two efficient algorithms were proposed by invoking the fixed-point iteration method and the manifold optimization method for the passive beamforming design. It was demonstrated that the proposed algorithms can achieve a higher performance and consume a lower complexity than the SDR approach. To solve the same problem, a branch-and-bound algorithm was further proposed by Yu {\em et al.}~\cite{Yu_optimal}, which is capable of obtaining a globally optimal solution. Though suffering from an extremely high computational complexity, the proposed branch-and-bound algorithm serves as a performance benchmark to verify the effectiveness of existing suboptimal algorithms. In~\cite{Ning137}, Ning {\em et al.} focused on an RIS-enhanced downlink MIMO system to maximize the SE. The passive beamforming was designed by using the sum of gains maximization principle and by utilizing the alternating direction method of multipliers. In~\cite{Ying_GMD}, Ying {\em et al.} considered an RIS-enhanced mmWave hybrid MIMO system, where the phase shifts at the RIS were designed by leveraging the angle information of the LoS BS-RIS channel. Moreover, Perovi{\'c} {\em et al.}~\cite{perovic2019channel} investigated RIS-assisted indoor mmWave communications, where two schemes were developed to maximize the channel capacity. Zhang {\em et al.}~\cite{Zhang_Capacity} characterized the fundamental capacity limit of RIS-aided point-to-point MIMO communication systems, by jointly optimizing the RIS reflection coefficients and the MIMO transmit covariance matrix. The communication capacity was maximized in both the narrowband transmission with frequency-flat fading channels and the broadband orthogonal frequency division multiplexing (OFDM) transmission over frequency-selective fading channels. Yang {\em et al.}~\cite{Yang_114} proposed a practical transmission protocol by considering the channel estimation for an RIS-enhanced OFDM system under frequency-selective channels. To reduce the required training overhead, the RIS reflection elements were divided into multiple groups and only the combined channel of each group has to be estimated. Based on the proposed grouping scheme, the achievable rate was maximized by jointly optimizing the power allocation at the transmitter and the phase shifts at the RIS with the AO method. In~\cite{You_113}, You {\em et al.} designed a transmission protocol by considering the channel estimation with discrete phase shifts at the RIS. To reduce the channel estimation errors, a low complexity discrete fourier transform (DFT)-Hadamard based reflection pattern scheme was developed. The achievable data rate was further maximized based on the estimated channel by designing the RIS phase shifts using the proposed successive refinement algorithm.

\item \emph{Sum rate maximization:} In~\cite{Huang_117}, Huang {\em et al.} maximized the sum rate in RIS-enhanced multi-user MISO downlink communications. By employing the zero-forcing precoding at the BS, the RIS reflection matrix and the power allocation were alternately optimized with the aid of the majorization-minimization approach. Moreover, the weighted sum rate maximization problem was investigated by Guo {\em et al.}~\cite{guo2019weighted}. Under the AO framework, the transmit beamforming was obtained using the fractional programming method, and three iterative algorithms were designed for optimizing the reflection coefficients in terms of different types of RIS reflection elements. In~\cite{Jung_96}, the asymptotic optimal discrete passive beamforming solution was derived and a modulation scheme was proposed to maximize the achievable sum rate for the RIS-enhanced multi-user MISO transmission. To further enhance the performance, Jung {\em et al.}~\cite{Jung_96} designed a joint user scheduling and transmit power control scheme, which can strike the tradeoff between the rate fairness and the maximum sum rate among the users. Mu {\em et al.}~\cite{Mu_124} focused their attention on the sum rate maximization problem in an RIS-enhanced MISO NOMA system with both ideal and non-ideal assumptions of RIS elements. The non-convex rank-one constraint of the passive beamforming design was handled by invoking the sequential rank-one constraint relaxation approach, which is guaranteed to obtain a locally optimal rank-one solution. Instead of optimizing the passive beamforming with the instantaneous CSI, Zhao {\em et al.} proposed a two-timescale transmission protocol for maximizing the achievable average sum rate in an RIS-enhanced multi-user system~\cite{Zhao_TwoTime}. To reduce the channel training overhead and complexity, the RIS phase shifts were firstly optimized with the statistical CSI. Then, the transmit beamforming was designed with the instantaneous CSI and the optimized RIS phase shifts.

\item \emph{User fairness:} Nadeem {\em et al.}~\cite{Nadeem_128} maximized the minimum SINR of an RIS-enhanced MISO system, where the BS-RIS-user link was assumed to be a LoS channel. A deterministic approximations was developed for the minimum SINR performance under the optimal linear precoder by employing the random matrix theory. As a result, the RIS phase shifts can be optimized using the channel's large-scale statistics, which can significantly reduce the overhead of the signal exchange~\cite{Nadeem_128}. Yang {\em et al.}~\cite{Yang_125} investigated the max-min rate problem in an RIS-enhanced NOMA system in both single-antenna and multi-antenna cases. A combined-channel-strength based user ordering scheme was proposed for achieving a near-optimal performance.
\end{itemize}
\begin{table*}[htbp]\scriptsize
\caption{Contributions on joint transmit and passive beamforming design}
\begin{center}
\centering
\resizebox{\textwidth}{!}{
\begin{tabular}{|l|l|l|l|l|l|}
\hline
\centering
 \textbf{Ref.} &\textbf{Scenarios} & \textbf{Phase shifts} & \textbf{CSI} & \textbf{Main Objectives} & \textbf{Techniques/Characteristics} \\
\hline
\centering
 \cite{wu2019intelligent} & SU/MU-DL-MISO & Continuous & Perfect & Transmit power &  AO-SDR-based algorithm and two stage algorithm\\
\hline
\centering
 \cite{Wu_138} & SU/MU-DL-MISO & Discrete & Perfect & Transmit power &  Near-optimal ZF-based successive refinement method\\
\hline
\centering
 \cite{han2019intelligent} & MU-DL-MISO & Continuous  & Perfect & Transmit power & Physical-layer broadcasting\\
\hline
\centering
 \cite{Fu_142} & MU-DL-MISO NOMA & Continuous  & Perfect & Transmit power & Alternating DC method\\
\hline
\centering
\cite{Zhu_NOMA} &  MU-DL-MISO NOMA & Continuous  & Perfect & Transmit power & Improved quasi-degradation condition\\
\hline
\centering
\cite{Zheng_NOMA} &  MU-DL-SISO NOMA & Discrete  & Perfect & Transmit power & Asymmetric and symmetric user pairing\\
\hline
\centering
\cite{huang2019reconfigurable} &  MU-DL-MISO & Continuous & Perfect & EE & RIS power consumption model\\
\hline
\centering
\cite{Zhou_141} & MU-DL-MISO & Continuous & Imperfect & Transmit power & The worst-case robust beamforming design\\
\hline
\centering
\cite{Zhou_IP} & MU-DL-MISO & Continuous & Imperfect & Transmit power & Imperfect cascaded channels at the transmitter\\
\hline
\centering
\cite{zappone2020overhead} & SU-DL-MIMO & Continuous & Estimated & EE & Overhead model for channel estimation and RIS configuration\\
\hline
\centering
\cite{Yu140} & SU-DL-MISO & Continuous  & Perfect & SE & Fixed point iteration and manifold optimization methods\\
\hline
\centering
\cite{Yu_optimal} & SU-DL-MISO & Continuous  & Perfect & SE & Branch-and-bound algorithm\\
\hline
\centering
\cite{Ning137} &  SU-DL-MIMO & Continuous  & Perfect & SE & Sum of gains principle\\
\hline
\centering
\cite{Ying_GMD} &  SU-DL-MIMO mmWave & Continuous & Perfect & SE & Broadband hybrid beamforming\\
\hline
\centering
\cite{perovic2019channel} & SU-DL-MIMO mmWave & Continuous & Perfect & Channel capacity &  RIS-assisted indoor mmWave environments\\
\hline
\centering
\cite{Zhang_Capacity} & SU-DL-MIMO & Continuous & Perfect & Channel capacity &  Frequency-flat fading and frequency-flat selective channels\\
\hline
\centering
\cite{Yang_114} & SU-DL-SISO OFDMA & Continuous & Estimated & Achievable rate & RIS element grouping scheme\\
\hline
\centering
\cite{You_113} & SU-UL-SISO & Discrete & Estimated & Achievable rate & Near-orthogonal DFT-Hadamard based reflection patterns \\
\hline
\centering
\cite{Huang_117} & MU-DL-MISO & Continuous  & Perfect & Sum rate  &  Majorization-minimization approach\\
\hline
\centering
\cite{guo2019weighted} & MU-DL-MISO & Continuous/Discrete & Perfect & Weighted sum rate & Iterative algorithms with closed-form expressions\\
\hline
\centering
\cite{Jung_96} & MU-DL-MISO & Discrete & Perfect & Sum rate & Interference-free modulation scheme\\
\hline
\centering
\cite{Mu_124} & MU-DL-MISO NOMA & Continuous/Discrete & Perfect & Sum rate & Sequential rank-one constraint relaxation approach\\
\hline
\centering
\cite{Zhao_TwoTime} & MU-DL-MISO & Discrete & Statistical CSI & Sum rate &  Low channel training overhead\\
\hline
\centering
\cite{Nadeem_128} & MU-DL-MISO & Continuous & Statistical CSI & Max-min SINR &  Signal exchange overhead reduction\\
\hline
\centering
\cite{Yang_125} & MU-DL-SISO/MISO NOMA & Continuous & Perfect & Max-min rate & Near-optimal NOMA user ordering scheme\\
\hline
\end{tabular}
}
\end{center}
\label{table:joint beamforming}
\end{table*}
All the aforementioned research contributions on the joint transmit and passive beamforming design are summarized in Table \ref{table:joint beamforming}. ``SU" and ``MU" represent single-user and multi-user, respectively. ``DL" and ``UL" represent downlink and
uplink, respectively.

\subsubsection{Approaches for passive beamforming design} An example of the joint transmit and passive beamforming design problem can be formulated as follows
\begin{subequations}\label{beamforming design}
\begin{align}
\mathop {\min /\max}\limits_{{\mathbf{w}},{\bm{\theta }}} &\;f\left( {{\mathbf{w}},{\bm{\theta }}|{\mathcal{H}}} \right)    \\
\label{transmit}{\rm{s.t.}}\;\;&{\mathbf{w}} \in {\mathcal{T}},\\
\label{passive}&{\bm{\theta }} \in {\mathcal{P}},
\end{align}
\end{subequations}
where ${\mathcal{H}}$ denotes the set of given CSI, ${\mathbf{w}}$ and ${\bm{\theta }}$ denote the transmit beamforming vector and passive beamforming vector, receptively, ${\mathcal{T}}$ and ${\mathcal{P}}$ denote the corresponding feasible set for ${\mathbf{w}}$ and ${\bm{\theta }}$, respectively. Here, $f\left( {{\mathbf{w}},{\bm{\theta }}|{\mathcal{H}}} \right)$ denotes the objective function that depends on ${\mathbf{w}}$ and ${\bm{\theta }}$ given the CSI. Let ${\theta _i} = {\beta _i}{e^{j{\phi _i}}}$ be the $i$th element of the passive beamforming vector ${\bm{\theta }}$. Depending on the specific implementation of the RIS, three case studies can be considered~\cite{Mu_124}.
\begin{itemize}
  \item \textbf{Continuous amplitude and phase shift}: In this case, it is assumed that the amplitude and phase shift of each RIS element can be adjusted continuously, which results in the following feasible set.
      \begin{align}\label{phi 1}
      {\mathcal{P}_1} = \left\{ {{\beta _i},{\phi _i}|{\beta _i} \in \left[ {0,1} \right],{\phi _i} \in \left[ {0,2\pi } \right)} \right\}.
      \end{align}
  \item \textbf{Constant amplitude and continuous phase shift}: In this case, it is assumed that the amplitude and phase shift of each RIS element are fixed, e.g., $\beta_i=1$, and can be adjusted continuously, respectively. The corresponding feasible set is given by
      \begin{align}\label{phi 2}
      {\mathcal{P}_2}= \left\{ {{\beta _i},{\phi _i}|{\beta _i} = 1,{\phi _i} \in \left[ {0,2\pi } \right)} \right\}.
      \end{align}
  \item \textbf{Constant amplitude and discrete phase shift}: In this case, it is assumed that the amplitude and phase shift of each RIS element are fixed, e.g., $\beta_i=1$, and can be adjusted based on a discrete set of values, respectively. The feasible set can be expressed as
      \begin{align}\label{phi 3}
      {\mathcal{P}_3} = \left\{ {{\beta _i},{\phi _i}|{\beta _i} = 1,{\phi _i} \in {{\mathcal{D}}}} \right\},
      \end{align}
      where ${\mathcal{D}} = \left\{ {0,\frac{{2\pi }}{N}, \cdots ,\left( {N - 1} \right)\frac{{2\pi }}{N}} \right\}$ and $N$ denotes the number of candidate phase shifts.
\end{itemize}
It is worth noting that the first two case studies are difficult to realize in practice. Due to the hardware constraints in fact, it is quite challenging to realize continuous amplitude and phase shift control. However, the first two case studies can be used to characterize the theoretical performance upper bounds of RISs.

By inspection of the three case studies just considered, we evince that the joint beamforming optimization problem is generally a non-convex problem since ${\mathbf{w}}$ and ${\bm{\theta }}$ are coupled together. The existing algorithms for solving the non-convex joint beamforming optimization in RIS-assisted wireless networks are mainly based on the AO method. The advantage of this approach is that, given the passive beamforming vector, the transmit beamforming design becomes a conventional problem, which has been extensively investigated. However, the passive beamforming design under given transmit beamforming vectors is still a non-trivial task to tackle. The main challenges to solve the problem include the unit modulus constraint and the discrete nature of the feasible set. In the following, we list the approaches employed in current research contributions for optimizing the passive beamforming. Table \ref{table:approach} summarizes the characteristics of those approaches.
\begin{table*}[htbp]\scriptsize
\caption{Summary of approaches to passive beamforming design}
\begin{center}
\centering
\begin{tabular}{|l|l|l|l|l|}
\hline
\centering
 \textbf{Approaches} & \textbf{Phase shifts} & \textbf{Advantages} &\textbf{Disadvantages} & \textbf{Ref.} \\
\hline
\centering
SDR & Continuous & Relax to convex problem & Require rank-one solution construction & \cite{wu2019intelligent,Zhu_NOMA}, etc.  \\
\hline
\centering
Quantization method & Discrete & Easy to implement & Substantial performance loss& \cite{Yang_125}  \\
\hline
\centering
Branch-and-bound & Continuous/Discrete & Optimal solution & Relatively high complexity & \cite{Wu_138,Yu_optimal}  \\
\hline
\centering
Iterative algorithms & Continuous/Discrete & Good complexity-performance tradeoff  & Performance depends on initialization  & \cite{Wu_138,Fu_142}, etc. \\
\hline
\end{tabular}
\end{center}
\label{table:approach}
\end{table*}
\begin{itemize}
  \item \textbf{SDR:} A common method for handling the non-convex unit-modulus constraint is to transform the passive beamforming vector into a rank-one and positive semi-definite matrix. By applying the SDR approach, which ignores the non-convex rank-one constraint, the original non-convex problem becomes a convex SDP problem that can be solved by using many efficient convex optimization tools. If the obtained matrix solution is not rank-one, Gaussian randomization methods are usually used. However, the constructed rank-one solution is, in general, suboptimal and it may be even infeasible for the original passive beamforming design problem, which not only causes some performance degradation but cannot guarantee the convergence of the AO-based iterative algorithm either.
  \item \textbf{Quantization method:} Under the assumption of finite resolution phase shifts, one method is to relax each discrete phase shift variable ${\phi _i}  \in \left\{ {0,\frac{{2\pi }}{N}, \cdots ,\left( {N - 1} \right)\frac{{2\pi }}{N}} \right\}$ into a continuous variable ${\phi _i}  \in \left[ {0,2\pi } \right)$. After solving the relaxed problem, the obtained continuous solutions are quantized to their nearest discrete values. However, the quantization method may lead to a substantial performance loss, especially for low-resolution phase shifts. Additionally, it is worth mentioning that the non-convex unit-modulus constraints still exist after applying the continuous relaxations.
  \item \textbf{Branch-and-bound:} Due to the non-convex nature of the passive beamforming design problem, it is challenging to obtain the optimal solution with standard convex optimization techniques. The branch-and-bound approach has been applied for solving polynomial-time (NP)-hard discrete and combinatorial optimization problems and some specific continuous optimization problems. For example, the branch-and-bound approach was adopted for deriving the optimal solution of the discrete passive beamforming design~\cite{Wu_138} and the continuous passive beamforming design~\cite{Yu_optimal}.
  \item \textbf{Iterative algorithms:} The main idea of iterative algorithms is to obtain a locally optimal or a high-quality suboptimal solution for the original problem at an acceptable computational complexity. Some iterative algorithms were developed for the passive beamforming design, such as the successive refinement algorithm~\cite{Wu_138,Zhang_Capacity,You_113,guo2019weighted}, the alternating DC algorithm~\cite{Fu_142}, the conjugate gradient search~\cite{huang2019reconfigurable}, the fixed point iteration method and the manifold optimization method~\cite{Yu140}, and the sequential rank-one constraint relaxation approach~\cite{Mu_124}. It was shown that these proposed iterative algorithms can achieve a good tradeoff between performance and computational complexity.
\end{itemize}
\subsection{Resource Management in RIS-enhanced Networks}
\begin{figure}[t!]
    \begin{center}
        \includegraphics[width=3.5in]{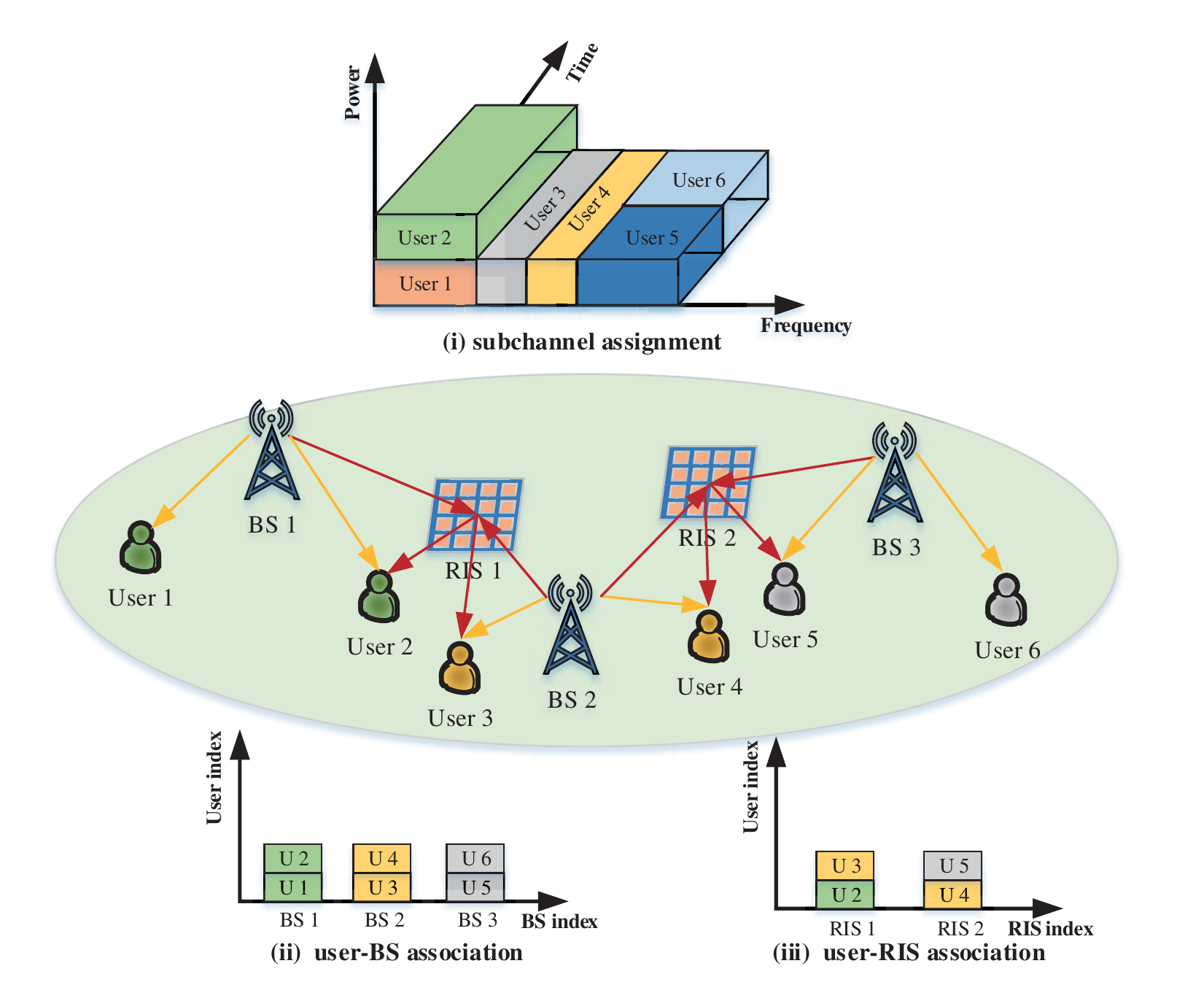}
        \setlength{\abovecaptionskip}{-0cm}
        \caption{Illustration of resource management in large-scale RIS-assisted networks.}
        \label{resource-management}
    \end{center}
\end{figure}
\subsubsection{Resource allocation problems} In Fig. \ref{resource-management}, a large-scale RIS-assisted transmission scenario is considered, where multiple BSs serve multiple users with the aid of multiple RISs. In this context, several key issues need to be discussed.
\begin{itemize}
  \item \emph{Subchannel assignment:} The bandwidth efficiency can be improved by properly allocating users to different subchannels. If the RIS elements are not frequency-selective, one single common RIS reflection matrix needs to be shared among all subchannels, which makes the resulting optimization problems challenging to solve. To address this difficulty, Yang {\em et al.}~\cite{Yang_129} proposed a dynamic passive beamforming scheme, where the resource blocks are dynamically allocated to different user groups with varied RIS phase shifts over different time slots. In~\cite{Zuo_NOMA}, Zuo {\em et al.} investigated the joint subchannel assignment, power allocation, and passive beamforming design problem in a multi-channel downlink RIS-NOMA network.
  \item \emph{User-RIS association:} In multi-RIS assisted multi-user communications, how to associate the users to different RISs is an interesting problem. The user-RIS association schemes in general determine the overall network performance. Considering a multi-RIS assisted massive MIMO system, Li {\em et al.}~\cite{Li_126} found that the automatic interference cancelation property holds for RISs with infinitely large sizes. Then, the considered max-min SINR problem can be transformed into a user-RIS association problem, which was efficiently solved by the proposed greedy search algorithm.
  \item \emph{Multi-cell RIS association:} In multi-cell scenarios, the optimization problem becomes much sophisticated for jointly considering the user-BS association, the user-RIS association, and the subchannel assignment. In some initial works considering multi-cell scenarios~\cite{pan2019Intelligent,Xie_multi}, RISs were deployed to enhance the performance of cell-edge users for simultaneously improving the received desired signal power and mitigating the received interference from other cells.
\end{itemize}
\subsubsection{Approaches to resource allocation problems} Scheduling different users with different subchannels/RISs/BSs is an NP-hard problem. Though the optimal solution can be obtained by exhaustively searching over all possible association combinations, it requires a prohibitively high computational complexity, especially for the large-scale networks in Fig. \ref{resource-management}. Therefore, low complexity and efficient algorithms have to be developed for striking a performance-versus-complexity tradeoff. In the following, we present some promising approaches, and discuss their advantages and disadvantages. Table \ref{table:approach RA} summarizes the characteristics of those approaches.
\begin{table*}[htbp]\scriptsize
\caption{Summary of approaches to resource allocation problems}
\begin{center}
\centering
\begin{tabular}{|l|l|l|l|}
\hline
\centering
 \textbf{Approaches}  & \textbf{Advantages} &\textbf{Disadvantages} & \textbf{Ref.} \\
\hline
\centering
Binary relaxation & Relax to convex feasible set & Existence of performance gap & -  \\
\hline
\centering
Matching theory &  Achieve near-optimal performance & Require predefined preference list& \cite{Zuo_NOMA}  \\
\hline
\centering
Heuristic algorithms &  Flexible complexity-performance tradeoff & Unstable performance & \cite{Li_126}  \\
\hline
\end{tabular}
\end{center}
\label{table:approach RA}
\end{table*}
\begin{itemize}
  \item \textbf{Binary relaxation:} One idea is to relax the binary variable $\alpha  \in \left\{ {0,1} \right\}$ into the continuous variable $\alpha  \in \left[ {0,1} \right]$, where $\alpha $ represents the user association state. By doing so, the non-convex integer constraint is relaxed to be convex, and conventional convex optimization techniques can be applied for solving the relaxed problem. It is worth pointing out that the relaxed problem might still be non-convex especially when the optimization variables are highly-coupled. Additional efforts, such as utilizing the successive convex approximation (SCA) method, are required to obtain an approximate solution. Moreover, this kind of relaxation may result in a considerable performance loss between the original integer problem and the relaxed one.
  \item \textbf{Matching theory:} Matching theory is a powerful method developed for solving the combinatorial user association problems. The user association combinational optimization problem in RIS-enhanced networks can be modeled as a high dimensional Users-BSs-RISs-Subchannels matching problem. Though high dimensional matching is NP-hard, it can be decomposed into several 2D matching subproblems, which can be efficiently solved. For example, Zuo {\em et al.}~\cite{Zuo_NOMA} applied many-to-one matching theory for the subchannel assignment in an RIS-enhanced NOMA system, which can achieve a near-optimal performance. However, leveraging matching theory requires to establish a predefined preference list for both users and resources. As the channel conditions always fluctuate in RIS-enhanced networks, these preference lists may need to be dynamically updated, which needs further investigations.
  \item \textbf{Heuristic algorithms:} For solving computationally complex problems, one commonly employed method is to develop heuristic algorithms, where approximate solutions for the original optimization problem can be obtained with an acceptable computational complexity. In the aforementioned works, the greedy search based heuristic algorithm was designed for solving the user-RIS association problem~\cite{Li_126}. However, the performance of heuristic algorithms is sensitive to the designed strategies, which is not always stable.
\end{itemize}

\subsection{Discussions and Outlook}
With the growing number of research contributions on RIS-enhanced communications, the advantages of RISs have been verified in terms of SE, EE, and user fairness. However, most of the existing treatises solved the non-convex joint beamforming optimization problem employing the AO method, which decouples the joint transmit and passive beamforming design into two subproblems. Though a high-quality suboptimal solution can be obtained, advanced optimization techniques are required for solving this problem to characterize the optimal performance gain introduced by RISs, which also provides an important benchmark for verifying the optimality of any other low complexity suboptimal algorithms. Furthermore, in RIS-enhanced communication systems, it is known that obtaining accurate CSI is rather challenging due to the nearly passive working mode of RISs. Investigating the robust joint beamforming design and resource allocation~\cite{Zhou_141,Zhou_IP} constitutes an important research direction for practical RIS implementations.

Besides passive beamforming, RISs also introduce the following DoFs, which can be further exploited to reap the benefits of RISs in the future work.
\indent
\begin{itemize}
  \item \emph{RIS deployment design:} The reflection link via the RIS experiences a severer path loss than the direct link. Therefore, the deployment location of the RIS has to be carefully designed to achieve considerable performance enhancements. How to jointly optimize the passive beamforming and the deployment location at the RIS as well as the wireless resource allocation at the AP/BS is a non-trivial task, which deserves further research efforts. In particular, one of prominent challenges is that the deployment location of the RIS determines both the path loss and the LoS components of the reflection channels, which causes the optimization variables to be highly-coupled. Therefore, efficient algorithms need to be designed. An initial study~\cite{Mu_deployment} has investigated the optimal RIS deployment strategy for both NOMA and OMA transmissions, which showed that asymmetric and symmetric RIS deployment locations among users are preferable for NOMA and OMA, respectively. Additionally, the RIS is usually deployed to avoid signal blockage to achieve a LoS dominated channel, thus having a small path loss. However, such a LoS channel based RIS deployment strategy may be ineffective, especially for RIS-assisted multi-user communications. This is because the LoS dominated channels are low-rank, the resulting ill-condition channel matrices significantly limit the achievable capacity even with a relatively small path loss. How to deploy the RIS to strike a balance between the path losses and the Non-LoS (NLoS) components of channels is another interesting problem, which deserves further research interests.
  \item \emph{Dynamical RIS configuration:} In \cite{Karasik_Joint,Mu_Capacity}, the authors revealed that, from an  information-theoretic perspective, the capacity-achieving transmission schemes need to dynamically adjusting the RIS. However, most of the existing research contributions assumed that the RIS reflection coefficients can be only adjusted once for each channel coherence duration. Note that one of the most typical application scenarios of RISs is to assist transmission of the users, who are static or moving slowly in the vicinity of RISs. In this case, the duration of one channel coherence block is usually tens of milliseconds, which is much larger than the time duration for adjusting the RIS (e.g., 220 microseconds in~\cite{arun2020rfocus}). Therefore, adjusting RIS multiple times in one channel coherence duration, namely the dynamic RIS configuration, is practically valid. This unique characteristic opens up new research opportunities, such as dynamical passive beamforming and resource allocation schemes, which merit further investigations.
\end{itemize}

\section{Machine Learning for RIS-enhanced Communication Systems}

ML techniques have gained remarkable interests in wireless communications due to their learning capability and large search-space~\cite{gacanin2020wireless,Xiao2020Artificial,wang2020artificial}. We survey existing research contributions, which apply ML techniques for tackling challenges in RIS-enhanced wireless networks. Finally, potential research challenges and opportunities of ML-empowered RIS systems are presented.

\subsection{Motivations and Architecture for Integrating ML in RIS-enhanced Wireless Networks}

In this subsection, we first present the challenges of the conventional RIS-enhanced wireless networks and the motivations for integrating ML in these networks, followed by the system architecture of ML-empowered RIS-enhanced wireless networks.

To effectively exploit RISs for optimizing wireless networks, preliminary research contributions have studied a number of technical challenges that include channel estimation/modeling, joint transmit and passive beamforming design, as well as resource allocation from the BS to the users. Powerful optimization techniques, such as convex optimization~\cite{guo2019weighted}, iterative algorithm~\cite{wu2019intelligent}, gradient descent approach~\cite{huang2019reconfigurable}, and alternating optimization algorithm~\cite{shen2019secrecy} have been adopted for addressing the aforementioned fundamental challenges. Although important insights have been gained by these research contributions. the following limitations still exist in conventional RIS-enhanced wireless networks:

\begin{itemize}
\item The users are generally assumed to be static for simplicity, i.e., the dynamic mobility of users is typically ignored. Another limitation in the existing literature is that the communication environment is assumed to be perfectly known, the differentiation of users' demand is always ignored as well.
\item The RIS/BS are not capable of learning from the unknown environment or from the limited feedback of the users. In practical applications of RISs in wireless networks, the system parameters are treated as random variables, which naturally leads itself to the derivation of insightful joint probability distributions conditioned on the users' tele-traffic demand and mobility. However, this is a highly dynamic stochastic environment, which is difficult for employing conventional optimization approaches. Additionally, the feedback from the users is usually resource-hungry and limited, which aggravates the challenges for the conventional RIS-enhanced wireless networks.
\item Finally, instantaneous CSI of all the channels are assumed to be available at the BS. However, CSI acquisition in RIS-enhanced wireless networks becomes more challenging than that in the conventional relay systems due to the passive nature of RISs, which also aggravates the challenge imposed on the conventional RIS-enhanced wireless networks.
\end{itemize}

\subsection{Deep Learning for RIS-enhanced Communication Systems}

Deep learning (DL) has shown great potentials to revolutionize communication systems. It can be applied in diverse areas of RIS-enhanced wireless networks due to its powerful learning capabilities~\cite{zappone2019model,Qin2019Deep,zappone2019wireless}.

The acquisition of timely and accurate CSI plays a pivotal role in wireless systems, especially in MIMO networks. However, CSI acquisition becomes more challenging due to the large number of antennas in massive MIMO systems~\cite{wen2018deep}. In order to tackle this challenge, a number of research contributions have adopted DL for estimating the CSI, especially for exploiting CSI structures beyond linear correlations.

In contrast to the conventional AF relay-aided wireless networks, in RIS-enhanced wireless networks, the RIS is a passive device, which is not capable of performing active transmission/reception and signal processing~\cite{chen2019channel}. In an effort to estimate a large number of unknown parameters caused by RISs, Taha {\em et al.}~\cite{taha2019enabling} exploited the DL method for learning the RIS reflection matrices directly from the sampled channel knowledge without any knowledge of the RIS array geometry. Liu {\em et al.}~\cite{liu2020deep} proposed a deep denoising neural network assisted compressive channel estimation for RIS-assisted mmWave systems with a low training overhead. Elbir {\em et al.}~\cite{elbir2020deep} presented a DL framework for channel estimation in the RIS-enhanced MIMO system. It was shown that the proposed convolutional neural networks
(CNNs)-based approach achieves lower normalized mean-square-error (NMSE) and more robust performance than other benchmarks.

The data-driven DL approach has the advantage of model-free representation or function learning such that no explicit models of the complicated wireless channels are needed, at the expense of requiring large amounts of training data and corresponding computational power. Thus, the DL method can be adopted for estimating the CSI of RIS-enhanced wireless networks.

Apart from the aforementioned applications of DL in RIS-enhanced wireless networks, Huang {\em et al.}~\cite{huang2019indoor} leveraged a deep neural network (DNN)-based approach in the indoor communication environment for estimating the mapping between a user's position and the configuration of the RIS to maximize the received SNR. Additionally, DL can also be applied for learning the optimal RIS phase shift configuration. Gao {\em et al.}~\cite{gao2020unsupervised} proposed a DL-based algorithm for optimally designing the phase shift of the RIS by training the DL offline. It can be observed that the proposed unsupervised learning mechanism outperforms the conventional optimization approach in terms of computational complexity. Khan {\em et al.}~\cite{khan2019deep} investigated the signal estimation and detection in the RIS-enhanced wireless networks. A DL-based approach was proposed for estimating channels and phase angles from a reflected signal received by an RIS. With the aid of DL, the bit-error-rate (BER) performance of the system was improved.

\subsection{Reinforcement Learning for RIS-enhanced Communication Systems}

Reinforcement learning (RL) is a powerful AI paradigm that can be used to empower agents by interacting with the environment. More explicitly, by exploiting the learning capability (e.g., learning from the environment, learning from the feedback of users, and learning from its mistakes) of the RL model, the challenges encountered in the conventional RIS-enhanced wireless networks may be mitigated, thus leading to improved performance.

The core idea of employing RL techniques in the RIS-enhanced wireless networks is that they allow the BS/RIS to improve their service quality by learning from the environment, from their historical experience, and from the feedback of the users~\cite{liu2019trajectory}. More explicitly, RL models can be used for supporting the BS/RIS (agents) in their interactions with the environment (states), whilst finding the optimal behavior (actions) of the BS/RIS. Furthermore, the RL model can incorporate farsighted system evolution (long-term benefits) instead of only focusing on current states. Thus, it is applied for solving challenging problems in the RIS-enhanced wireless networks.

\begin{figure*}[t!]
    \begin{center}
        \includegraphics[width=16cm]{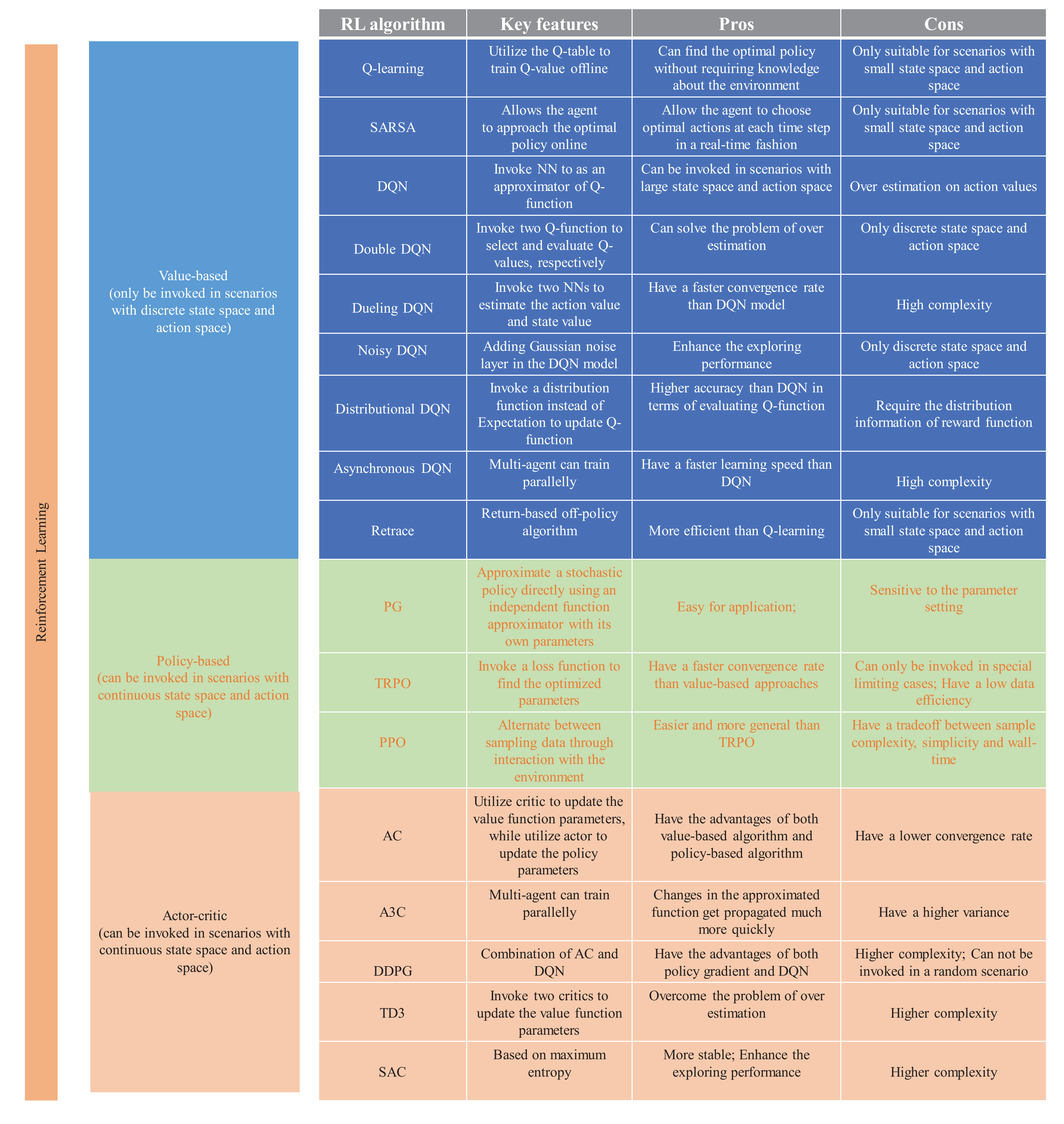}
     \caption{Key features, pros and cons of RL algorithms~\cite{liu2019trajectory,van2016deep,wang2015dueling,fortunato2017noisy,dabney2018distributional,mnih2016asynchronous,schulman2017proximal,hamalainen2018ppo,lillicrap2015continuous,haarnoja2018soft,luong2019applications}. (PG represents Policy Gradient, TRPO denotes Trust Region Policy Optimization, PPO represents Proximal Policy Optimization, AC denotes Actor-Critic, A3C represents Asynchronous Advantage Actor-Critic, DDPG denotes Deep Deterministic Policy Gradient, TD3 represents Twin Delayed DDPG, SAC denotes Soft Actor-Critic)}
            \label{RL}
    \end{center}
\end{figure*}

As illustrated in Fig.~\ref{RL}, the RL algorithms can be divided into three categories, namely, value-based algorithms, policy-based algorithms, and actor-critic algorithms. Both advantages and disadvantages exist in the RL algorithms. Since RISs have discrete phase shifts, the DQN algorithm is more suitable for tackling the corresponding phase shift design problem.

To fully reap the benefits of deploying RISs in wireless networks, the joint transmit and passive beamforming design of the RIS-enhanced system has been considered in MISO systems~\cite{huang2020reconfigurable,feng2020deep}, OFDM-based systems~\cite{taha2020deep}, wireless security systems~\cite{Helin2020Deep} and millimeter wave systems~\cite{zhang2020millimeter} with the aid of RL algorithms. In contrast to the AO method, which alternately optimizes the transmit beamforming at the BS and the passive beamforming at the RIS, the RL-based solution is capable of simultaneously designing them. More explicitly, Huang {\em et al.}~\cite{huang2020reconfigurable} applied a deep deterministic policy gradient (DDPG) based algorithm for maximizing the throughput by utilizing the sum rate as instant rewards for training the DDPG model. In the proposed model, the continuous transmit beamforming and RIS phase shift were jointly optimized with low complexity. Taha {\em et al.}~\cite{taha2020deep} proposed a deep reinforcement learning (DRL) based algorithm for maximizing the achievable communication rate by directly optimizing interaction matrices from the sampled channel knowledge. In the proposed DRL model, only one beam was utilized for each training episode. Thus, the training overhead was avoided, while the dataset collection phase was not required. Zhang {\em et al.}~\cite{zhang2020millimeter} presented a DRL based algorithm for maximizing the throughput with both perfect and imperfect CSI. A quantile regression method was applied for modeling a return distribution for each state-action pair, which modeled the intrinsic randomness in the MDP interaction between the RIS and communication environment. Helin {\em et al.}~\cite{Helin2020Deep} considered the application of RISs to PLS. The system secrecy rate was maximized with the aid of the DRL model by jointly optimizing the beamforming and phase shift matrices under different users' QoS requirements and time-varying channel conditions. Additionally, post-decision state and prioritized experience replay schemes were utilized to enhance the learning efficiency and secrecy performance.

\subsection{A Novel Architecture of ML-empowered RIS-enhanced Wireless Networks}

\begin{figure*}[t!]
    \begin{center}
        \includegraphics[width=16cm]{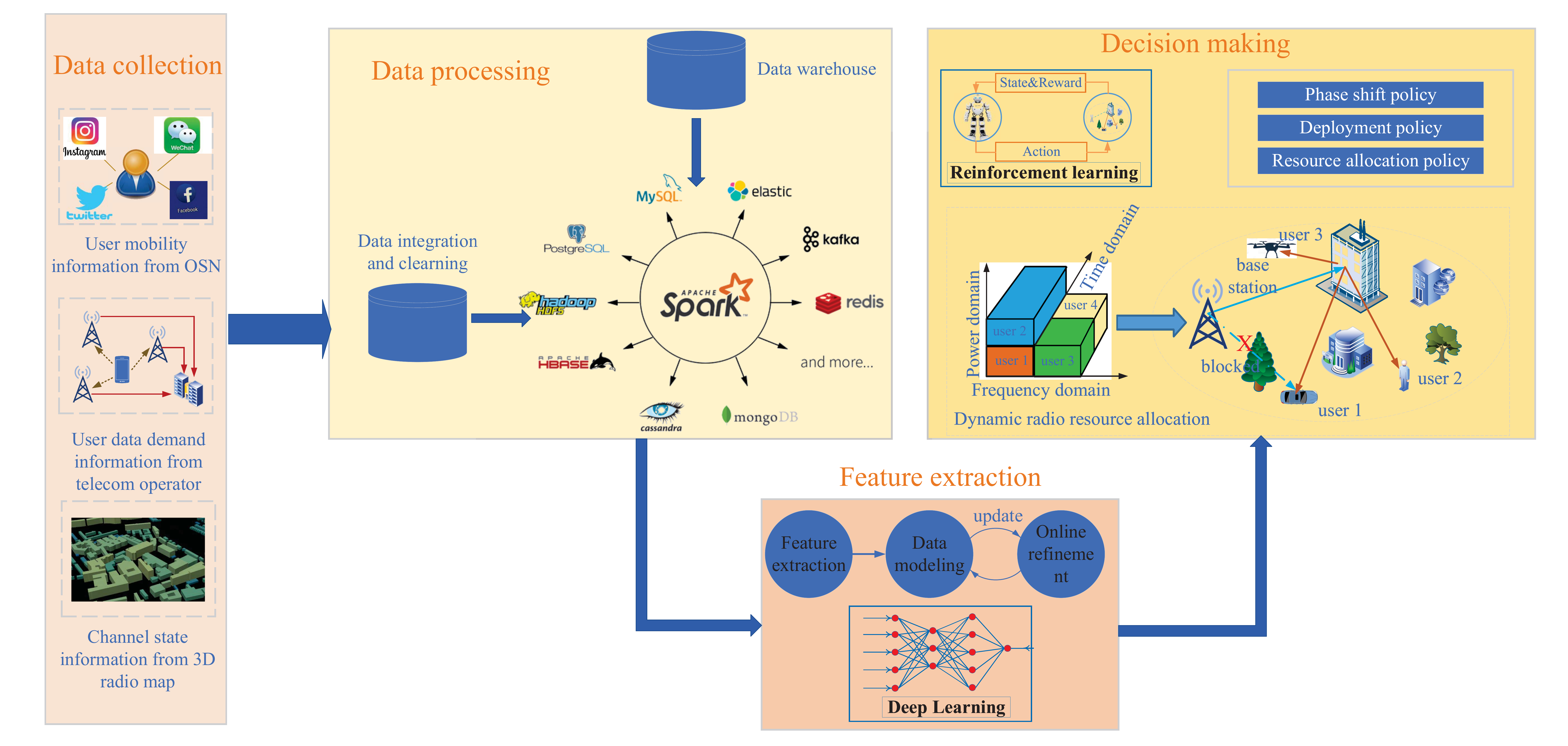}
     \caption{Architecture of ML-empowered RIS-enhanced wireless networks.}
            \label{architecture}
    \end{center}
\end{figure*}

As a benefit of the ML-based framework, many challenges in conventional wireless communication networks have been circumvented, leading to enhanced network performance, improved reliability, and agile adaptivity~\cite{Xiao2020Artificial}. Fig.~\ref{architecture} illustrates a novel ML-empowered architecture for RIS-enhanced wireless networks. As shown in this figure, RISs are installed on the facade of a building for enhancing the wireless performance~\cite{di2019smart,Qingqing2020Towards}. The RIS is linked with a controller, which controls the reflecting elements for hosting the functionality of phase-shifting and amplitude absorption. A two-step approach is applied in the proposed ML-empowered RIS-enhanced wireless networks.

\begin{itemize}
\item As illustrated in the data collection, data processing, and feature extraction parts in Fig.~\ref{architecture}. The associated user information (e.g., device type, position, data rate demand, mobility, caching demand, and computing ability) is collected, stored, and processed. Thus, the users' behaviors and requirements can be predicted for efficiently deploying and operating the RIS. Meanwhile, the predicted information can be modified online with the currently collected data as the input.
\item Given the extracted features, adaptive schemes are leveraged for controlling the RISs, designing the phase shifts, resource allocation, and interference cancellation.
\end{itemize}

In the proposed ML-empowered RIS-enhanced wireless networks, RISs are capable of rapidly adapting to the dynamic environment by learning both from the environment and from the feedback of the users.

Research on the RIS deployment is fundamental and essential. However, there is a paucity of research on the problem of RIS position determination. Additionally, current research contributions mainly consider the performance optimization for both single-user and multi-user scenarios by optimizing the phase shift and/or precoding solutions of the RIS-enhanced communication systems~\cite{Nadeem_128,ye2019joint,liang2019large,jung2019performance,pan2019Intelligent}.

Considering the RIS deployments based on the users' mobility information and particular data demand implicitly assumes that the long-term movement information and tele-traffic requirement of users are capable of being learned/predicted. With this proviso, the deployment and control method of RISs may be designed periodically for maximizing the long-term benefits and hence reducing the control overhead. By considering the long-term mobility and data demand of users, RIS-enhanced wireless networks become highly dynamic systems. Meanwhile, in an effort to maximize the service quality in an unknown environment, RISs are supposed to learn by interacting with the environment and adapting the control/deployment policy based on the limited feedback of the users to overcome the uncertainty of the environment.

\begin{figure*}[t!]
    \begin{center}
        \includegraphics[width=14cm]{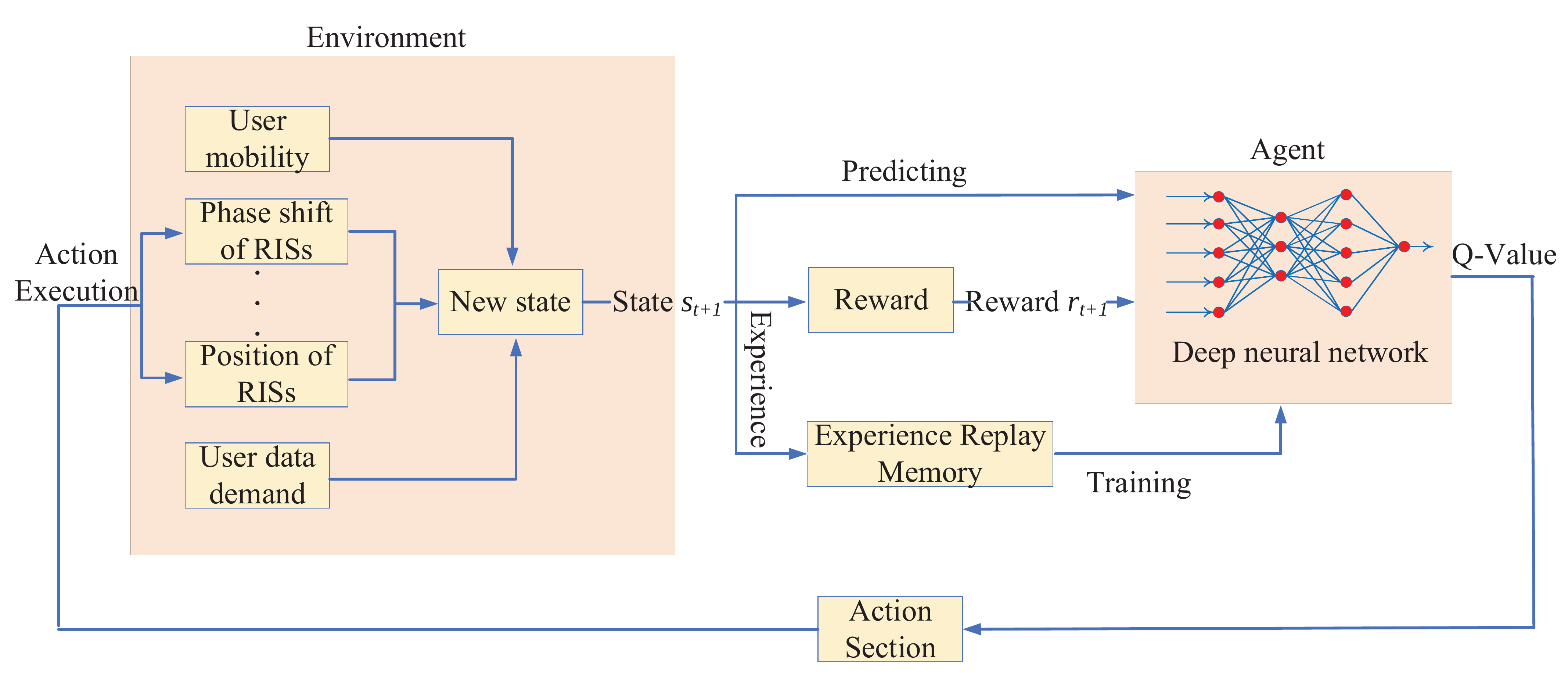}
     \caption{DRL model for RIS networks.}
            \label{DQN}
    \end{center}
\end{figure*}

In this subsection, an RL-based model is presented to jointly design the deployment policy and phase shift policy of RISs while considering the time-varying data demand of users. As illustrated in Fig.~\ref{DQN}, in the RL-based model, the BS acts as an agent. Since a controller is installed, the BS can control both resource allocation policy for users and the RIS's position and phase shifts. At each timeslot, the BS periodically observes the state of the RIS-enhanced system. The state space consists of the RIS phase shifts, the allocated power to each user, as well as coordinates of both the RIS and users. An action is carried out by the BS for selecting the optimal control policy. The actions contain changing positions and varying phase shifts of the RIS, as well as varying the allocated power. The key underlying principle of the decision policy is carrying out an action that makes the DQN model obtain the maximum Q-value at each time slot. Following each action, the BS receives a penalty/reward, $r_t$, determined by the formulated objective function.

\begin{itemize}
\item \textbf{State of the RL model.} The state space consists of four parts: 1) the current phase shift of each reflecting element at the RIS; 2) the current 3D position of the RIS; 3) the current 2D position of each user; 4) the current power allocated from the BS to each user.
\item \textbf{Action of the RL model.} The action space consists of three parts: 1) the variable quantity of the $n$-th reflecting element's phase shift; 2) the moving direction and distance of the RIS; 3) the variable quantity of the $k$-th user's transmit power.
\item \textbf{Reward of the RL model.} The reward function is decided by the EE of the system. When action taken by the BS improves EE, the BS obtains a reward. Otherwise, when a reduction occurs in EE, the BS receives a penalty.
\end{itemize}

\begin{figure} [t!]
\centering
\includegraphics[width=3.3in]{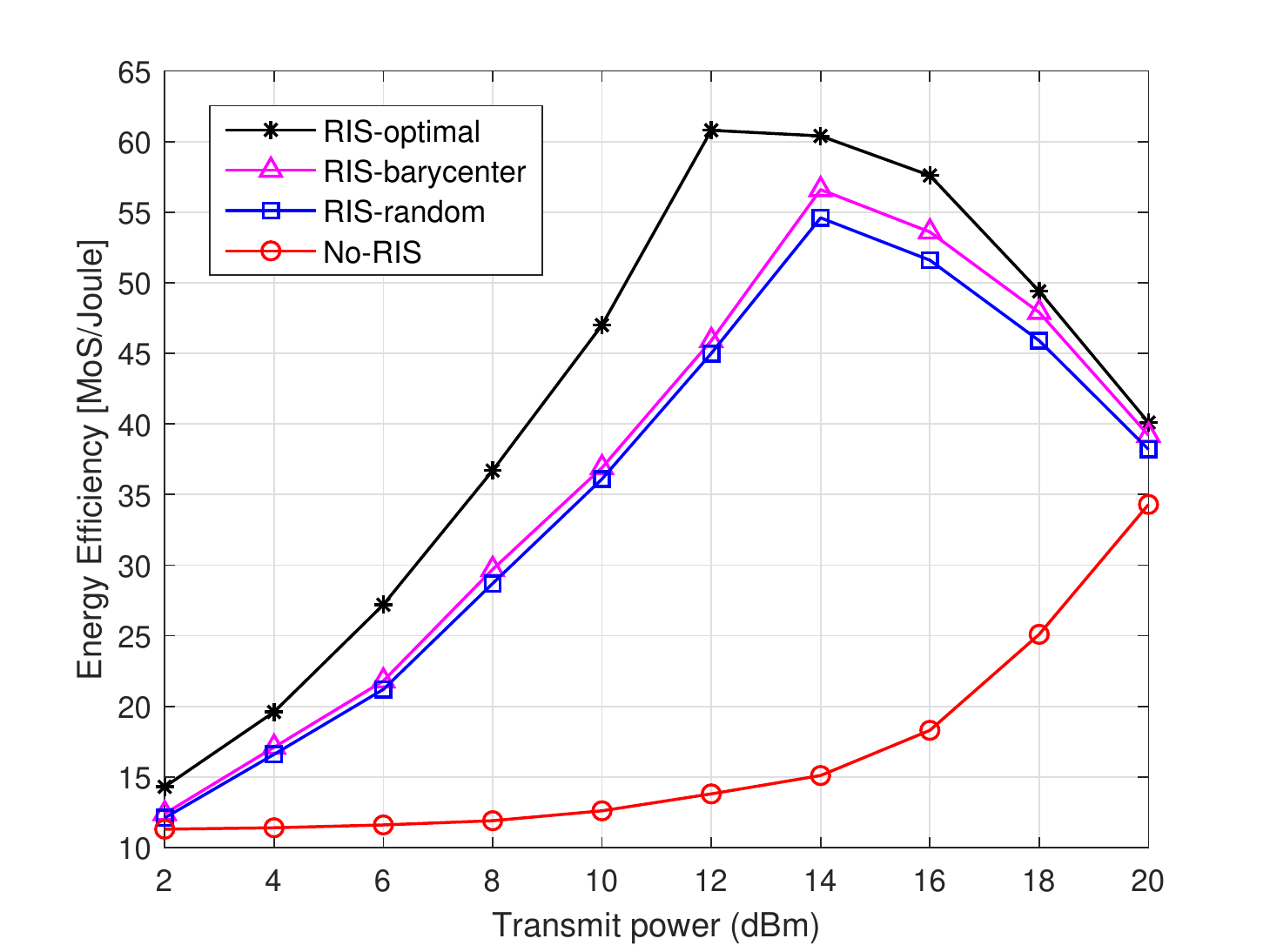}
\caption{EE with and without RIS~\cite{liu2020ris}.}\label{EEvsnoIRS}
\end{figure}

Fig.~\ref{EEvsnoIRS} characterizes the EE of the system in networks both with and without the assistance of an RIS. The EE is defined as the ratio between the system achievable sum mean opinion score (MOS) and the sum energy dissipation in Joule. It was shown that the EE of the system is enhanced by employing an RIS. The RIS-barycenter line indicates that the RIS is placed at the barycenter of all users. The RIS-random line indicates that the RIS is randomly deployed, while the RIS-optimal line indicates that the RIS is deployed at the optimal position derived from the proposed decaying double deep Q-network (${{\text{D}}^{\text{3}}}{\text{QN}}$) algorithm. The results of Fig.~\ref{EEvsnoIRS} confirm that there exists an optimal position for the RIS as far as the EE of the RIS-enhanced system is concerned. The performance of the RIS-enhanced system is improved by deploying the RIS at the optimal position compared to the random deployment strategy and the strategy of placing it at the barycenter.

\subsection{Other ML Techniques for RIS-enhanced Communication Systems}

Besides DL and RL techniques, a range of supervised learning and unsupervised learning algorithms have been applied in the current generation wireless networks. Thus, these approaches can also be adopted for tackling challenges in the RIS-enhanced wireless systems.

\subsubsection{Supervised Learning techniques for RIS-enhanced Communication Systems}

As one of the key branches of ML, powerful supervised learning techniques, such as regression, decision tree and random forest, K-nearest neighbors (KNN), support vector machines (SVM), and Bayes classification, have been adopted in diverse scenarios for tackling challenges such as spectrum sensing~\cite{umebayashi2017efficient}, traffic/QoE prediction~\cite{feng2017proactive}, channel/antenna selection~\cite{thilina2015dccc}, and networking association~\cite{abouzar2011action}. In the RIS-enhanced wireless networks, supervised learning algorithms can also be applied for solving the related problems with sufficient training data due to their advantages of low complexity and fast convergence speed.

\subsubsection{Unsupervised Learning techniques for RIS-enhanced Communication Systems}

In contrast to the supervised learning techniques, unsupervised learning methods do not rely on prior knowledge, which is not data-hungry. Thus, the unsupervised learning algorithms~\cite{wang2019thirty} such as K-means clustering, expectation-maximization, principal component analysis (PCA), and independent component analysis (ICA) can be applied in the RIS-enhanced wireless networks for tackling challenges such as BS deployment, user clustering/association~\cite{liu2019reinforcement}, channel/network state detection~\cite{assra2015approach}, data aggregation~\cite{morell2016data}, and interference cancellation~\cite{li2017digital}.

\subsubsection{Federated Learning techniques for RIS-enhanced Communication Systems}

Federated learning, which explores training statistical models directly on remote devices, has become a focal point in the area of large-scale ML and distributed optimization~\cite{niknam2019federated}. Since the federated learning algorithm is trained at the edge in distributed networks, the inaccessibility of private data is no longer a problem. Due to its privacy-preserving nature, federated learning algorithms can be applied for the deployment and design of multiple RISs, where each RIS can act as a distributed learner, trains its generated data and transfers its local model parameters instead of the raw training dataset to an aggregating unit. Thus, the deployment and design policy can be learned in a decentralized manner.

\subsection{Discussions and Outlook}

By exploiting ML learning capabilities, the aforementioned challenges encountered in RIS-enhanced wireless networks may be mitigated. This is due to the reason that RISs can learn by interacting with the environment and adapt the control/deployment policy based on the feedback of users to overcome the dynamic/uncertainty of the environment. It can be learned that the RL model can incorporate farsighted system evolution instead of only focusing on current states, which can reap long-term benefits for RIS-enhanced wireless networks. However, ML models will also pose some new challenges, including the layer design for the DL model, the state-action construction, and the reward function design for the RL model. In addition, simultaneously employing multiple RISs becomes more challenging due to the cooperation amongst RISs. Hence, intelligent deployment and design for multi-RIS enhanced wireless networks is highly desired. Finally, in the current ML models, either discrete or continuous state space is modeled for optimizing parameters in RIS-enhanced wireless networks, while joint discrete and continuous parameters exist in the networks. Hence, the joint discrete and continuous state space design in ML-enabled RIS-enhanced wireless networks is still challenging and constitutes an interesting topic.

\section{Integrating RISs with Other Technologies Towards 6G}

Current research contributions have proved that RIS-enhanced wireless networks are capable of obtaining tuned channel gains, improved QoS, enhanced coverage range, and reduced energy dissipation. These significant performance enhancements can be applied to diverse wireless communication networks. In this section, we identify the major issues and research opportunities on the path to 6G associated with the integration of RISs and other emerging technologies, such as NOMA, PLS, SWIPT, UAV-enabled wireless networks, and autonomous driving networks.

\subsection{NOMA and RIS}

In an effort to improve the SE and user connectivity of RIS-enhanced wireless networks, power-domain NOMA technology is adopted, whose key idea is to superimpose the signals of two users at different powers for exploiting the spectrum more efficiently by opportunistically exploring the users' different channel conditions~\cite{Liu_physical_scurity_NOMA,liu2017non}. Li {\em et al.}~\cite{li2019joint} considered a MISO-NOMA downlink communication network for minimizing the total transmit power by jointly designing the transmit precoding vectors and the reflecting coefficient vector. In~\cite{Yang_114}, Yang {\em et al.} jointly optimized the phase shifts matrix of the RIS, as well as the power allocation from the BS to the users. Thus, the minimum decoding SINR of all users was maximized for optimizing the throughput of the system by considering user fairness. Ding {\em et al.}~\cite{RIS_zhiguo_simple} proposed a novel design of RIS assisted NOMA networks. It can be observed in~\cite{RIS_zhiguo_simple} that, the directions of users' channel vectors can be aligned with the aid of the RIS, which emphasizes the importance of implementing NOMA technology. For an RIS-NOMA system, the core challenge is that the decoding order is dynamically changed due to the configuration of phase shifts of the RIS. Mu {\em et al.}~\cite{Mu_124} proposed an RIS-enhanced multiple-antenna NOMA transmission framework to maximize the throughput of the system by considering the NOMA SIC decoding order condition. The SCA technique and sequential rank-one constraint relaxation based algorithm were applied to obtain a locally optimal solution. Ni {\em et al.}~\cite{ni2020resource} proposed a resource allocation framework in multi-cell RIS-NOMA networks, where the achievable sum rate was maximized by solving the joint optimization problem of user association, sub-channel assignment, power allocation, phase shifts design, and decoding order determination.

In contrast to the conventional MIMO-NOMA systems, RIS-NOMA technology can overcome the challenges of the dynamic environment such as random fluctuation of wireless channels, blocking, and user mobility in an energy-efficient manner. The NOMA system can obtain tuned channel gains, improved fair resource allocation, enhanced coverage range, and high EE with the aid of RISs~\cite{sousa2020role}. However, NOMA also gives rise to new challenges when integrated with RISs. For multi-antenna NOMA transmission, the decoding order is not determined by the users' channel gains order, since additional decoding rate conditions need to be satisfied to guarantee successful SIC~\cite{Mu_124}. Additionally, both the active beamforming and passive phase shift design affect the decoding order among users and user clustering, which makes the decoding order design, user clustering, and joint beamforming design highly-coupled in RIS-NOMA networks.

\subsection{PLS and RIS}
It has been shown that RISs are capable of simultaneously enhancing the desired signal power at the intended user and mitigating the interference power at other unintended users~\cite{wu2019intelligent}. Inspired by this result, several researchers explored the potential performance gain in the context of PLS by applying the RIS~\cite{Yu_Secure,Cui_149,shen2019secrecy,Chu_Secure,Dong_Secure,Chen_150,Guan_AN,Yu_151}. Yu {\em et al.}~\cite{Yu_Secure} considered an RIS-enhanced multiple-input single-output single eavesdropper (MISOSE) channel, where the eavesdropper is equipped with a single antenna. The secrecy rate was maximized by jointly optimizing the transmit beamforming and the RIS phase shift matrix by using an AO-based algorithm. It was demonstrated that the secrecy performance can be significantly improved by deploying the RIS. Cui {\em et al.}~\cite{Cui_149} focused on the scenario where the eavesdropper has a better direct channel condition than that of the legitimate receiver and they are also highly correlated in space, where the achievable secrecy rate is rather limited in conventional communications. However, it was shown that the direct signals and the reflected signals can be destructively combined at the eavesdropper with the aid of RISs, thus significantly improving the secrecy rate. The same problem was further investigated in~\cite{shen2019secrecy,Dong_Secure} by considering a multi-antenna eavesdropper or legitimate receiver. Chu {\em et al.}~\cite{Chu_Secure} minimized the transmit power while satisfying the secrecy rate requirement in the RIS-enhanced MISOSE system. Chen {\em et al.}~\cite{Chen_150} studied the minimum secrecy rate maximization problem in the RIS-enhanced multi-user multiple-input single-output multiple eavesdropper (MISOME) system by considering both the continuous and discrete RIS phase shifts. Injecting artificial noise (AN) is an effective technique to enhance the secrecy rate~\cite{AN}. Motivated by this result, Guan {\em et al.}~\cite{Guan_AN} examined the effectiveness of employing AN in an RIS-enhanced MISOME system. The achievable secrecy rate was maximized by jointly optimizing the transmit beamforming, the passive beamforming, and AN. The results verified the necessity of using AN, especially for systems with a large number of eavesdroppers. Yu {\em et al.}~\cite{Yu_151} considered an RIS-enhanced multi-user MISOME system under imperfect CSI with the aim of maximizing the sum rate, subject to the maximum information leakage constraint. An efficient AO-based algorithm was developed to optimize the transmit beamforming, the AN covariance matrix, and the RIS phase shifts. Numerical results showed that significant secrecy performance gains can be achieved by the RIS.

One critical issue of the RIS-enhanced PLS is that the joint design of transmit and passive beamforming requires the CSI of both AP-eavesdropper and RIS-eavesdropper links, which is quite challenging to obtain. This is because besides the nearly passive working mode of RISs, in practice, eavesdroppers usually stay almost silent to hide their positions and only detect signals in the air. Therefore, robust joint beamforming designs under the imperfect CSI of the eavesdropper are essential to guarantee secure transmission. Moreover, given the uncertainty of eavesdroppers, deploying the RIS may increase the probability of information leakage since the eavesdropper can receive not only the direct signals from the AP but also the reflected signals from the RIS. The situation may become even worse when there are multiple cooperative eavesdroppers. In this case, setting a protected zone to establish an eavesdropper-exclusion area with carefully deployed RISs would help to enhance the secrecy performance, which deserve further investigations.

\subsection{SWIPT and RIS}
SWIPT is an attractive technique for future IoT networks. However, the low EE at the energy receivers is the main bottleneck in practical SWIPT systems. To overcome this limitation, deploying the RIS is a promising solution and the RIS-assisted SWIPT has been investigated in~\cite{Wu_SWIPT_letter,Tang_SWIPT,Pan_SWIPT,Wu_SWIPT}. In~\cite{Wu_SWIPT_letter}, Wu {\em et al.} investigated an RIS-assisted SWIPT system, subject to individual SINR requirements of information receivers. The weighted sum power received by energy receivers was maximized by jointly optimizing the transmit and passive beamforming with the proposed AO-based algorithm. Moreover, Tang {\em et al.}~\cite{Tang_SWIPT} maximized the minimum received power among energy receivers. The results in~\cite{Wu_SWIPT_letter} and~\cite{Tang_SWIPT} showed that deploying an RIS can improve the energy harvesting efficiency. Pan {\em et al.}~\cite{Pan_SWIPT} studied the weighted sum rate maximization problem in the RIS-assisted SWIPT MIMO system, subject to the energy harvesting requirement of each energy receiver. A block coordinate descent (BCD)-based algorithm was designed to find a Karush-Kuhn-Tucker (KKT) stationary point of the original optimization problem. Wu {\em et al.}~\cite{Wu_SWIPT} extended the RIS-assisted SWIPT system into a multi-RIS case, where the transmit power was minimized while satisfying the different QoS constraints at information users and energy users. It was shown that the RIS enlarges the wireless power transfer range and reduces the number of required energy beams.

Note that the above research contributions studied performance gain of deploying RISs for SWIPT mainly from the communication perspective and ignored the EM characteristic of RISs. As discussed in previous sections, there are substantial differences between the near-field region and the far-field region of RISs. Therefore, sophisticated EM-based wireless power transfer models are required for fully reaping the benefits of RISs, which need to be investigated in future work.

\subsection{UAV and RIS}

RISs can be applied in UAV-enabled wireless networks, where UAVs are employed to complement and/or support the existing terrestrial cellular networks~\cite{wang2018joint,osseiran2014COM}. An RIS enhances the UAV coverage and service quality by compensating for the power loss over long distances, as well as forming virtual LoS links between UAVs and mobile users via passively reflecting their received signals. Li {\em et al.}~\cite{li2019reconfigurable} jointly optimized the UAV trajectory and the RIS phase shifts in an iterative manner. It was shown in~\cite{li2019reconfigurable} that, the average achievable rates of the users were significantly improved with the aid of RISs. On the other hand, RISs can also be applied in UAV-aided wireless relay networks for enhancing performance. Zhang {\em et al.}~\cite{zhang2019reflections} considered the effective placement of a single UAV, which was equipped with an RIS to assist the mmWave downlink transmission while considering user mobility. By jointly designing the UAV trajectory and the RIS reflection parameters, a virtual LoS connection between the BS and users was guaranteed. Thus, both the average data rate and the achievable downlink LoS probability were improved. Yang {\em et al.}~\cite{YangOn2020_UAV} derived the analytical expressions of outage probability, BER, and average capacity by approximating the PDF of the instantaneous SNR in RIS-assisted UAV relaying systems. Mu {\em et al.}~\cite{Mu_UAV} proposed a novel RIS-aided multi-UAV NOMA transmission framework, where an RIS was deployed to enhance the desired signal strength between UAVs with their served ground users while mitigating the inter-UAV interference. Liu {\em et al.}~\cite{Xiao2020RISUAV} integrated UAVs in RIS-enhanced wireless networks for enhancing the service quality of the UAV. With the aid of RISs, the energy consumption of the UAV was significantly reduced.

Due to the fact that UAVs are battery-powered, how to reduce their energy consumption is one of the key challenges. The limited flight-time of UAVs (usually under 30 minutes) hampers the wide commercial roll-out of UAV-aided networking. By deploying RISs, one can adjust the RIS phase shift instead of controlling the UAV movement for forming virtual LoS links between the UAV and the users. Therefore, the UAV can maintain hovering status only when the virtual LoS links can not be formed even with the aid of the RIS. By invoking the aforementioned protocol, the total energy consumption of the UAV is minimized, which in turn, maximizes the UAV endurance. Additionally, by mounting a compact distributed laser charging (DLC) receiver or wireless power transmission (WPT) receiver antenna inside the UAVs, while a DLC/WPT transmitter is deployed on the ground or the building roof, the UAVs can be charged as long as they are flying within the coverage range of the DLC/WPT transmitter~\cite{Liu2016Charging,liu2019trajectory}. However, the LoS connection between UAVs and the charging stations/vehicles have to be guaranteed, which is challenging in the urban scenario when the LoS link between UAVs and charging stations/vehicles are blocked by high-rise buildings with a high probability. RISs are capable of smartly reconfiguring the wireless propagation environment by forming virtual LoS links between UAVs and the charging stations/vehicles via passively reflecting their received signals. Thus, the quality of charging service is enhanced with the aid of the RIS.

The RIS-enhanced UAV communication scenario is naturally a highly dynamic one, which falls into the field of ML. When considering both the trajectory design of UAVs and the phase shift design of RISs, the former one can be formed as a continuous state space while the latter one is usually formed as a discrete one. Hence, how to simultaneously deal with both continuous and discrete state space is challenging in ML-empowered RIS-enhanced UAV networks.

\subsection{Autonomous Driving/Connected Vehicles and RIS}

RISs can also be deployed in vehicle-to-infrastructure (V2I) assisted autonomous driving systems, where V2I components are employed to complement the costly onboard units (OBUs). V2I networks enable autonomous vehicles (AVs) or connected vehicles (CVs) to receive reliable real-time traffic information from BSs, the information is collected by roadside base stations (RBSs) and transmitted from RBSs to BSs, which facilitates the interaction among AVs/CVs and road users, hence enhancing their safety and traffic efficiency~\cite{yao2018v2x,vivacqua2018self,liu2020enhancing,williams2018information}. Since AVs/CVs quality and reliability are non-negotiable, the AVs/CVs system must be real-time, while the transmission is supposed to be 100\% reliable. However, the service quality of current V2I communication systems cannot be guaranteed due to the complex channel terrain in the urban environment and the complexity of road conditions, such as bad weather. Makarfi {\em et al.}~\cite{Makarf2020Reconfigurable} and Wang~\cite{Wang2020Outage} proved that the performance of vehicular networks can be significantly improved with the aid of RISs. Since the RISs are made of EM material, which can be installed on key surfaces, such as building facades, highway polls, advertising panels, vehicle windows, and even pedestrians' clothes. With the massive deployment of RISs, a virtual LoS connection between the BSs and AVs, as well as between the RSUs and BSs will be guaranteed, which enhances the reliability of V2I communications.

In RIS-enhanced autonomous driving systems, the driving safety of AVs is the primary consideration. In terms of safety, collisions have to be avoided, while the traffic rules also need to obey. Additionally, in RIS-enhanced V2I-assisted autonomous driving systems, the wireless service quality for AVs has to be guaranteed at each timeslot. Hence, how to improve the reliability of RIS-enhanced autonomous driving systems is an open and challenging problem.

\subsection{Discussions and Outlook}

The studies of RISs have unveiled promising research opportunities, such as NOMA, PLS, SWIPT, UAV-enabled wireless networks, and autonomous driving networks. Recent research contributions have proved that RIS-enhanced wireless networks can achieve tuned channel gains, improved QoS, enhanced coverage range, and reduced energy dissipation. However, the network and beamforming designs are highly coupled due to dynamic control of the RIS phase shifts, which brings challenges to these new research directions.

\section{Conclusions, Challenges, and Potential Solutions}

\subsection{Concluding remarks}

In this paper, recent research works on RIS-enhanced wireless networks proposed for applications to next-generation networks have been surveyed with an emphasis on the following aspects: operating principles of RISs, performance evaluation of multi-antenna assisted RIS systems, joint beamforming design and resource allocation for RISs, ML in RIS-enhanced wireless networks, and their integration with other key 6G technologies. We have highlighted the advantages and limitations of employing RISs for communication applications. Further research efforts are needed to bridge the complex physical models of the different RISs implementations with widely used communication models. We have considered the performance evaluation of multi-antenna assisted RIS systems by systematically surveying existing designs for RIS-enhanced wireless networks from the views of performance analysis, information theory, and optimization. In addition, we have discussed existing research contributions that applying ML tools for tackling the dynamic essence of the wireless environment such as random fluctuations of wireless channels and user mobility. Design guidelines for ML-empowered RIS-enhanced wireless networks have also been discussed. The benefits of integrating RISs with NOMA, UAV-terrestrial networks, PLS, SWIPT, and AVs/CVs have been discussed. However, the research of RIS-enhanced wireless networks is still at a very early stage and there are ample opportunities for important contributions and advances in this field. Some of them are listed as follows.

\subsection{Challenges and Potential Solutions}

\subsubsection{CSI Acquisition}

The acquisition of timely and accurate CSI plays a pivotal role in RIS-enhanced wireless systems, especially in MIMO-RIS and MISO-RIS networks. The majority of current research contributions assume perfect CSI available at the BS, RISs controllers, as well as the users. However, obtaining CSI in RIS-enhanced wireless networks is a non-trivial task, which requires a non-negligible training overhead. Additionally, in RIS-assisted NOMA networks, users in each cluster have to share the CSI with each other for implementing SIC. However, due to the passive characteristic of RISs, the CSI acquisition and exchanging are non-trivial. Potential solutions can be developed by employing DL methods for exploiting CSI structures beyond linear correlations.

\subsubsection{Pareto-Optimization for Satisfying Multiple Objectives}

In contrast to the conventional wireless networks, RIS-enhanced wireless networks are characterized by more rapidly fluctuating network topologies and more vulnerable communication links. Furthermore, RISs are more likely to be deployed in an environment with heterogeneous mobility profiles. Hence, the networks operate in a complex time-variant hybrid environment, where the classic mathematical models have limited accuracy. Additionally, the challenging optimization problems encountered in RIS-enhanced wireless networks usually have to satisfy multiple objectives (e.g., delay, throughput, BER, and power) in order to arrive at an attractive solution. To elaborate, by definition it is only possible to improve any of the metrics considered at the cost of degrading at least one of the others. The collection of Pareto-optimal points is referred to as the Pareto front. However, determining the entire Pareto-front of optimal solutions is still challenging. Potential solutions may be investigating near-real-time ML-aided Pareto-optimization for tackling the high-dynamic adaptation of RIS-enhanced wireless networks.

\begin{spacing}{1.75}
\bibliographystyle{IEEEtran}
\bibliography{mybib}
\end{spacing}
\end{document}